\documentclass[structabstract]{aa}
\usepackage{txfonts}
\usepackage{graphicx} 
\usepackage[colorlinks=true,urlcolor=blue,citecolor=blue,linkcolor=blue]{hyperref}
\usepackage{natbib}
\bibpunct{(}{)}{;}{a}{}{,}

\def\kms {{\mathrm{km}\,\mathrm{s}^{-1}}}

\def\ha {H$\alpha$}
\def\hb {H$\beta$}

\defcitealias{Wedemeyer2009}{WR09}
\defcitealias{Barklem2007HNLTE}{PB07}
\defcitealias{Barklem2002}{PB02}

\begin{document}

\title{How realistic are solar model atmospheres?}
\titlerunning{{How realistic are solar model atmospheres?}}

\author{T. M. D. Pereira\inst{1, 2}, M. Asplund\inst{1,3}, R. Collet\inst{1,3}, I. Thaler\inst{3}, R. Trampedach\inst{4}, J. Leenaarts\inst{2}}
\authorrunning{Pereira et al.}

\institute{Research School of Astronomy and Astrophysics, Australian National University, Cotter Rd., Weston, ACT 2611, Australia
\and Institute for Theoretical Astrophysics, University of Oslo, PO Box 1029, Blindern, N--0315, Norway 
\and Max-Planck-Institut f\"ur Astrophysik, Postfach 1317, D--85741 Garching b. M\"unchen, Germany 
\and JILA, University of Colorado and National Institute of Standards and Technology, 440 UCB, Boulder, CO 80309, USA
}

\date{Received 2 February 2013 /
Accepted 17 April 2013}

\abstract 
{Recently, new solar model atmospheres have been developed to replace classical 1D LTE hydrostatic models and used to for example derive the solar chemical composition.}%
{We aim to test various models against key observational constraints. In particular, a 3D model used to derive the solar abundances, a 3D MHD model (with an imposed 10\,mT vertical magnetic field), 1D NLTE and LTE models from the PHOENIX project, the 1D MARCS model, and the 1D semi-empirical model of Holweger \& M\"uller.} 
{We confront the models with observational diagnostics of the temperature profile: continuum centre-to-limb variations, absolute continuum fluxes, and the wings of hydrogen lines. We also test the 3D models for the intensity distribution of the granulation and spectral line shapes.} 
{The predictions from the 3D model are in excellent agreement with the continuum centre-to-limb observations, performing even better than the Holweger \& M\"uller model (constructed largely to fulfil such observations). The predictions of the 1D theoretical models are worse, given their steeper temperature gradients.
 For the continuum fluxes, predictions for most models agree well with the observations. No model fits all hydrogen lines perfectly, but again the 3D model comes ahead. 
The 3D model also reproduces the observed continuum intensity fluctuations and spectral line shapes very well.} 
{The excellent agreement of the 3D model with the observables reinforces the view that its temperature structure is realistic. It outperforms the MHD simulation in all diagnostics, implying that recent claims for revised abundances based on MHD modelling are premature. Several weaknesses in the 1D hydrostatic models (theoretical and semi-empirical) are exposed. The differences between the PHOENIX LTE and NLTE models are small. We conclude that the 3D hydrodynamical model is superior to any of the tested 1D models, which gives further confidence in the solar abundance analyses based on it.}

\keywords{Sun:~photosphere -- line:~formation -- stars:~atmospheres --
  Sun:abundances -- Sun:~granulation}

\maketitle

\section{Introduction}

Solar model atmospheres are a cornerstone in stellar astronomy. The Sun is a natural reference when studying other stars, and realistic solar photosphere models are essential to infer solar parameters such as its chemical composition. The wealth of solar data available can be used to rigorously test and constrain photosphere models. This testing provides invaluable insight about the model physics and its degree of realism,
paving the way for building realistic models of other stars.

With the significant increases in computational power of recent years, the classical approximations used when modelling stellar atmospheres have started to be challenged. Of these approximations, the most significant are the assumption of a static 1D atmosphere with a mixing length type treatment of convection, and the assumption of local thermodynamical equilibrium (LTE). Although at present no model of a stellar atmosphere is able to relax these two assumptions simultaneously, efforts have been made to tackle each of these approximations individually. On the geometry/convection side, realistic 3D hydrodynamical time-dependent simulations of convection have been developed and used as models of the solar photosphere \citep[\emph{e.g.}][]{SteinNordlund1998,Asplund2000,Freytag2002,Vogler2004,Carlsson2004,Caffau2008b}. On the radiative transfer side, \citet{ShortHauschildt2005} have computed a 1D non-LTE (NLTE) hydrostatic solar model atmosphere with the PHOENIX code \citep{HauschildtBaron1999} following the pioneering work in this regard by \citet{Anderson1989}. %

The application of 3D solar models to abundance analysis \citep[\emph{e.g.}][]{Asplund2004, Caffau:2010} has resulted in a revised solar photospheric metallicity of $Z=0.0134$ \citep{Asplund2009}, substantially smaller than previous canonical values (\emph{e.g.} $Z=0.0201$ in \citealt{AndersGrevesse1989} and $Z=0.0169$ in \citealt{GS98}). The realistic treatment of convection and velocity fields in the 3D models resulted in an excellent agreement between predicted and observed line shapes and bisectors, not possible with the 1D models even with the free parameters of micro- and macro-turbulence \citep{Asplund2000}. This agreement is a strong indicator of how realistic the model is. Additionally, when compared with observations of the solar granulation at high spatial resolution, the 3D models correctly predict the characteristic size and lifetimes of the granules \citep[\emph{e.g.}][]{SteinNordlund1998,Nordlund:2009}. However, the results are still controversial because by using a lower solar metallicity the previous excellent agreement between solar interior models and helioseismology deteriorates significantly \citep[\emph{e.g.}][]{Bahcall2005,Basu2008,Serenelli:2009}. 

A criticism of the 3D solar models sometimes raised is that while they have been tested against many spectral lines, they lack a thorough testing of temperature structure, such as the continuum centre-to-limb variation and absolute continuum fluxes \citep{Basu2008}. A correct temperature structure is of the utmost importance to abundance studies. Using a 1D horizontal and temporal average of the 3D model of \citet{Asplund2000}, \citet{Ayres2006} suggest that the 3D model fails to describe the observed centre-to-limb variation and its temperature gradient is too steep. The first claim is partly dismissed by \citet{Koesterke2008}, showing that the 1D average is not a valid approximation of the full 3D model for temperature profiling and that the performance of the 3D model used by \citet{Asplund2000} and \citet{AGS05} in the continuum centre-to-limb variation is comparable to that of theoretical 1D models -- a view corroborated by \citet{Pereira2008} and \citet{TrujilloShchukina2009}. However, \citet{Koesterke2008} agree with \citet{Ayres2006} in that the temperature gradient of this particular older 3D model is slightly too steep in the continuum forming layers. As mentioned by \citet{Asplund2009}, an improved radiative transfer treatment resulted in a new 3D model with a more realistic temperature gradient (less steep than before). %

The aim of this work is to systematically test the temperature structure of several models of the solar photosphere, including 3D and 1D NLTE models. We use the new 3D hydrodynamical model employed by \citet{Asplund2009}, the 10~mT 3D MHD model of Thaler et al. (in preparation) and recent 1D NLTE and LTE models from the PHOENIX project \citep{HauschildtBaron1999}. We also employ the widely used 1D MARCS model \citep{MARCS2008} and the semi-empirical 1D model of \citet{HM1974}.

\begin{figure*} 
  \centering
  \includegraphics[width=0.49\textwidth]{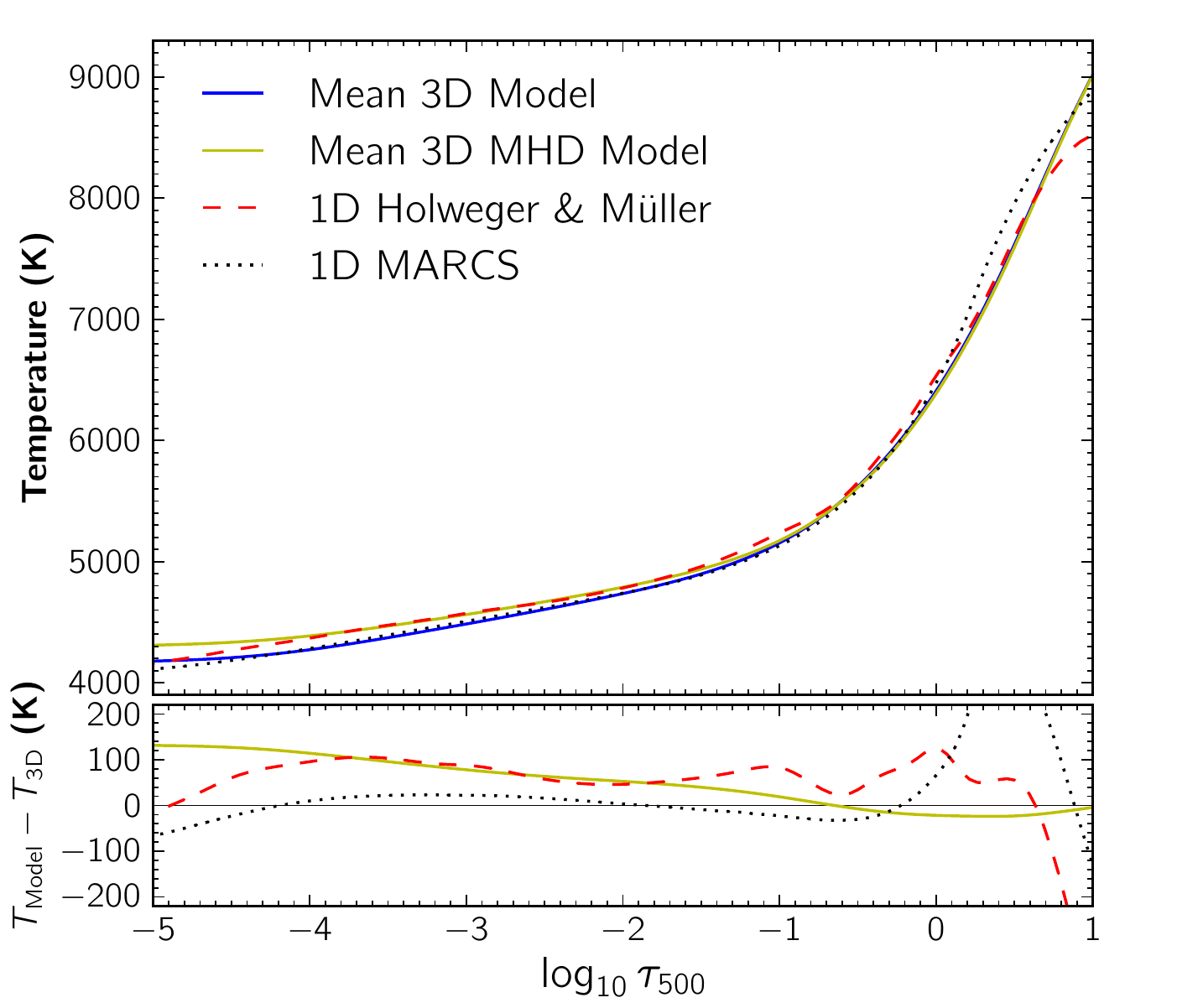}  
  \includegraphics[width=0.49\textwidth]{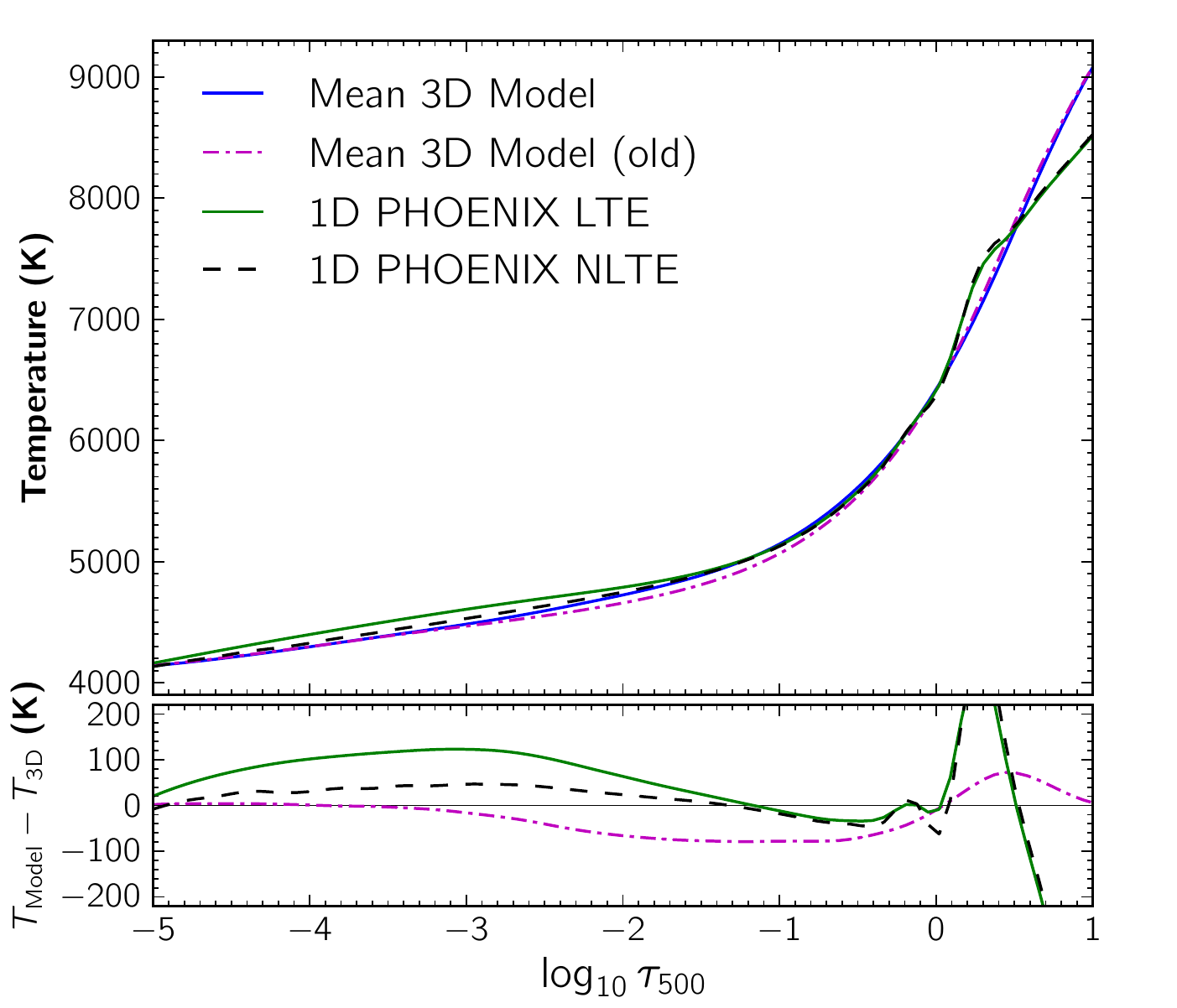}
  \caption{Temperature structure of the 3D and 1D models, plotted against the optical depth at 500~nm. For the 3D models structure represents the temporal and spatial mean (over $\tau_{500}$ iso-surfaces). \emph{Bottom panels:} differences between the 3D model and a given model (legend according to the top panels).}
  \label{fig:ttau}
\end{figure*}

The models used are described in more detail in Sect.~\ref{sec:models}. To compare their temperature structure with the observations we use three classical tests: the continuum centre-to-limb variation in Sect.~\ref{sec:clv}, the absolute continuum flux distribution in Sect.~\ref{sec:flux} and the wings of hydrogen lines in Sect.~\ref{sec:hlines}. In addition, we also test the 3D model against observations of the continuum intensity distribution and $\Delta I_\mathrm{rms}$ in Sect.~\ref{sec:contint}. Sect.~\ref{sec:felines} contains a comparison of the predicted and observed line shapes, including inferred abundances, bisectors and line shifts, for a sample of \ion{Fe}{i} and \ion{Fe}{ii} lines. Our conclusions are given in Sect.~\ref{sec:conc}. 

\section{Model atmospheres\label{sec:models}}

\subsection{3D Model}
We use the same 3D hydrodynamical model solar atmosphere that was adopted by \cite{Asplund2009} for the derivation of photospheric solar abundances.
The solar surface convection simulation was performed using the 3D, radiative, hydrodynamical, conservative,  {\sc stagger-code} \citep{NordlundGalsgaard1995}.
In the simulation, the equations for the conservation of mass, momentum, and energy are solved together with the radiative transfer equation for a representative volume of solar surface ($6{\times}6{\times}3.8$~Mm$^3$) on a Cartesian mesh with $240^3$ numerical resolution.
The horizontal grid is equidistant, while the vertical depth scale has a non-constant spacing optimised to better resolve the layers at the photospheric transition where temperature gradients are at their steepest.
Open, transmitting, boundaries are assumed at the top and bottom of the simulation domain, and periodic boundary conditions are enforced horizontally.
It is important for the lower boundary to be transmitting, to avoid homogenising the otherwise highly asymmetric (between up and down) convective flows.

The simulation domain completely covers the Rosseland optical depth range $-5 \leq \log \tau_\mathrm{Ross} \leq 7$.%
The radiative transfer equation is solved using a long-characteristics Feautrier-like scheme down to ${\tau}_\mathrm{Ross}{\approx}300$, and the diffusion approximation is employed in the deeper layers.
The main improvement over the simulation of \cite{Asplund2000} is the treatment of opacities and line-blocking in particular. 
In both the old (2005) and new (2009) 3D simulations, continuous and line opacities are included via a statistical method, called \emph{opacity binning} or \emph{multi-group method} \citep{Nordlund1982}: wavelengths are sorted into \emph{opacity bins} according to the strength of the opacity and the corresponding LTE source functions are added together within each bin.
In the original binning scheme, \cite{Nordlund1982} used the Rosseland average $\kappa_0$ of the opacities in the continuum bin, scaled by a constant factor for each of the other bins, $\kappa_\mathrm{j}\,=\,{\kappa_0}{10} ^{{j}{\Delta}{x}}$, in practice assuming ${\Delta}{x} = 1$ 
and $j\,=\,0,{\ldots},11$. 
The transition to free streaming for optical depths in the continuum bin ${\tau}_0{\ll}1$ is ensured by an exponential bridging (in $\tau_0$ ) to an intensity-weighted mean opacity. 
The multi-group method was further developed by \cite{Skartlien2000}, who relaxed the approximation of $\kappa_j$ just being a scaled $\kappa_0$, and instead computed the actual Rosseland average for each bin. This also ensures that the Rosseland mean of the bin-wise opacities converges to the actual Rosseland mean of the monochromatic opacities. 
The new 3D model has further been improved by sorting opacities into bins not only according to opacity strength but also according to wavelength, and allowing arbitrary bin sizes; a similar  binning criterion has been implemented by \citet{Caffau2008} in the $\mathrm{CO{^5}BOLD}$ code. 
For the present simulation, continuous opacities are taken from \citet[][ and subsequent updates]{Gustafsson1975} and line opacities from the latest MARCS stellar atmosphere package \citep{MARCS2008}. The solar chemical composition by \cite{AGS05} and the equation-of-state by \citet[][ and subsequent updates]{MHDI} are adopted.

The positions of the bin borders are then optimised with respect to a monochromatic radiative transfer calculation for the simulation's average temperature and density stratification taken on surfaces of constant Rosseland optical depth.
The generalisation of the bins and the optimisation reduce the differences between the radiative heating of the monochromatic and the binned solution by a factor of five, to within less than $1$\%. 
The effect of radiative heating and cooling on the simulation is therefore faithfully reproduced by this opacity-binning, for the average atmosphere. We have also performed a 3D monochromatic radiative transfer calculation on the full 3D simulation cube for a single snapshot, and found a $<5$~K ($0.08$\%) difference in $T_{\mathrm{eff}}$ between the monochromatic and the binned solution.
Before being subjected to scientific tests, the simulation has been fully relaxed, and has subsequently run for the $45$ solar minutes used here ($90$ snapshots). 
The relaxation process included extracting energy from radial $p$-modes, and ensuring that the total flux is statistically constant with depth, that no drifts are present in the thermodynamical quantities at the bottom boundary, and that the vertical grid's resolution is enough for the radiative transfer in the photosphere. 

For the calculations presented here the original simulation was interpolated to a coarser 50$\times$50$\times$82 resolution (with a finer vertical depth scale) to save computing time. The effective temperature of the 90 snapshots used is $T_{\rm eff} = 5778 \pm 2$\,K, close to the observed value of $5777 \pm 3$~K \citep{WillsonHudson1988}.

\subsection{3D MHD Model}

In addition to the 3D simulation, we also test a 3D magneto-hydrodynamical (MHD) simulation from Thaler et al. (in preparation). This simulation was performed using the same code and physical ingredients of the 3D hydrodynamical model of \citet{Asplund2009}. It uses \mbox{240$\times$240} grid points horizontally and 220 vertically. It
has the same horizontal size of \mbox{6$\times$6~Mm$^2$}
as the 3D hydrodynamical model, but a slightly
shorter vertical extension of 3.34~Mm. It extends 2.7~Mm into the
convection zone and reaches 0.645~Mm up the
photosphere. As in the 3D hydrodynamical model, the horizontal grid is
equidistant and the vertical depth scale optimised to resolve the
photosphere. The boundary conditions are the same as in the \citet{Asplund2009} simulation.
On a thermodynamically relaxed hydrodynamical snapshot, a
vertical magnetic field of 10 mT was overimposed.
The magnetic field is kept vertical at the bottom boundary and tends toward a potential field at the top.
This configuration was run for 120 min of solar time.
For the results presented here, a sequence of 38~min was used, taken after this simulation had run for 68~min.

The choice of 10~mT for B was made as it is a reasonable value for the quiet sun's mean field strength \citep[e.g.][]{Trujillo-Bueno:2006} and because it allows a comparison with the middle MHD model of \citet{Fabbian:2010,Fabbian:2012}.

\subsection{1D Models}

We use two types of 1D models: theoretical and semi-empirical. The semi-empirical 1D model of \citet{HM1974} was built from a range of observables to reproduce the mean physical quantities of the solar photosphere. Most importantly, it was constructed to follow the observed continuum centre-to-limb variation between 0.5--300\,$\mu$m and the line depths of $\approx$900 spectral lines. This is an important detail to note when considering our centre-to-limb variation comparison, although the observations we employ are more recent than the ones available when the Holweger \& M\"uller model was built. Historically, the Holweger \& M\"uller model has been the atmosphere of choice when deriving solar abundances. It assumes hydrostatic equilibrium and does not explicitly include convection. 

Of the 1D hydrostatic theoretical model atmospheres we include the LTE, line-blanketed solar MARCS model, which is a reference for the MARCS grid of model atmospheres \citep{MARCS2008}. We also include an LTE and an NLTE model from the PHOENIX project \citep{Hauschildt1999}. These two models have been computed for the solar abundances of \citet{AGS05} and the same input physics as for the PHOENIX Gaia grid \citep{PhoenixGAIA}. They differ only in their treatment of atomic level populations with the NLTE model having been computed with a NLTE treatment of H, He, C, N, O, Mg and Fe.

To ensure consistency when using our line formation code, opacities and equation of state, for the 1D models we took the $T(\tau)$ relation from their respective references and integrated $P_{\mathrm{gas}}$ in optical depth assuming hydrostatic equilibrium to obtain the pressures and densities that yield the same $T(\tau)$ when using our opacities and equation of state. We note that this is an often overlooked source of error when comparing results for supposedly the same model atmosphere since this pressure-integration is not always carried out. In fact the \citet{HM1974} model is essentially only an updated version of the \citet{Holweger1967} model with a new pressure-integration due to their effects on opacities and thermodynamics.

\begin{figure} 
  \centering
  \includegraphics[width=0.49\textwidth]{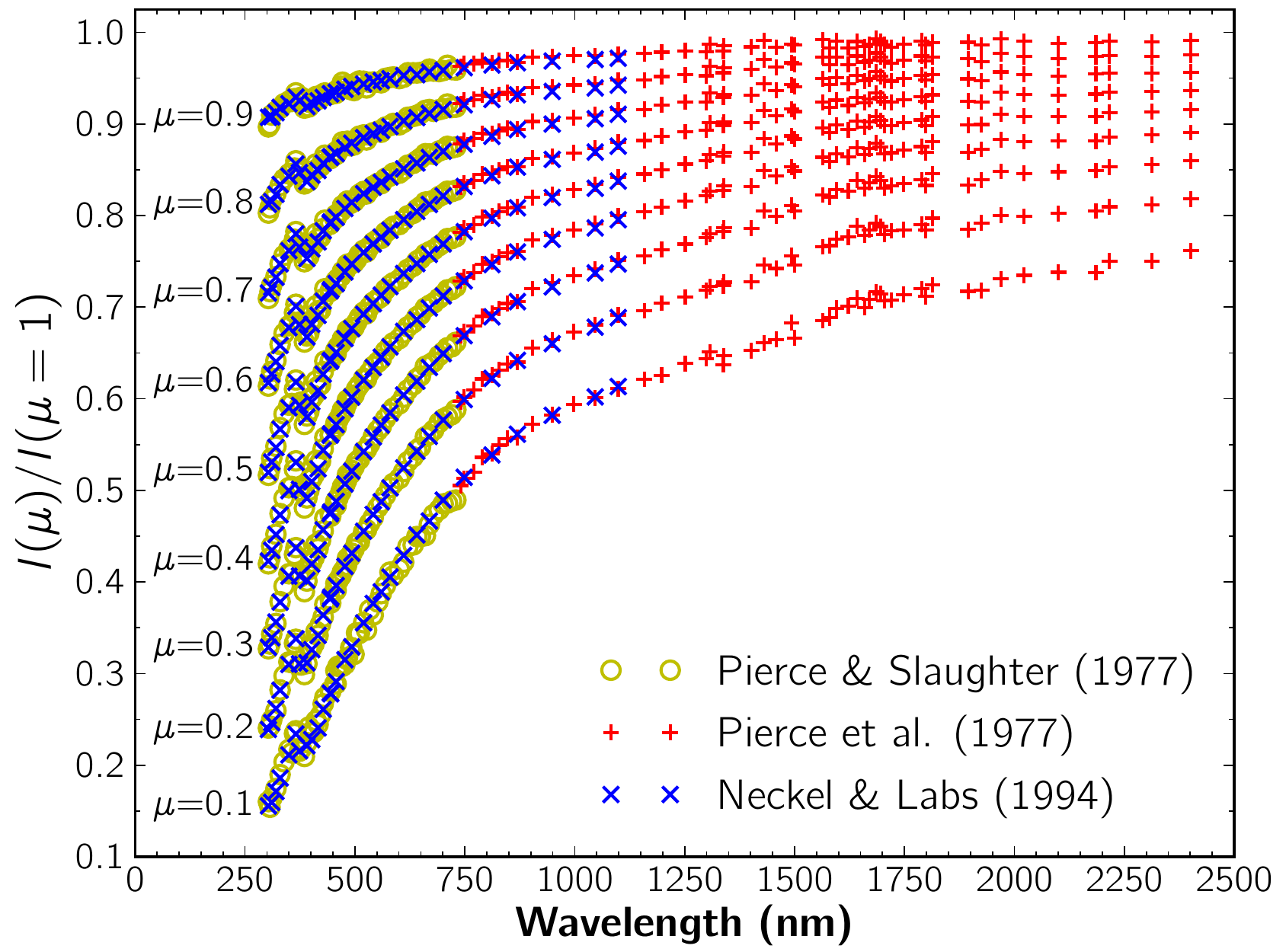}
  \caption{Continuum centre-to-limb observations in the visible and near-infrared, from \citet{Pierce1977a}, \citet{Pierce1977} and \citet{NeckelLabs1994}. In the wavelength region where the sets overlap we use only \citet{NeckelLabs1994} for our comparisons.}
  \label{fig:clv_obs}
\end{figure}

\subsection{Mean stratification}

In Fig.~\ref{fig:ttau} we compare the mean temperature structure of the 3D hydrodynamical model with other models. On the left, with the 3D MHD model of Thaler et al. (in preparation), the 1D Holweger \& M\"uller and MARCS models (left) and on the right with the old 3D model of \citet{Asplund2000}, and the PHOENIX LTE and NLTE models. The mean structure of the 3D models was calculated by averaging over $\tau_{500}$ iso-surfaces. 

The 3D model of \citet{Asplund2000} is provided here for a quick comparison. This older model was run with a precursor of the \textsc{stagger-code}, and was used to derive the solar abundances of \citet{AGS05}. The largest difference between the old and new models is the treatment of radiation. The new model has an improved multi-group opacity scheme with 12 bins, whereas the old model had a more approximate scheme and used only 4 bins. From Fig.~\ref{fig:ttau} one can see that the major consequence of the improved scheme is a warmer upper photosphere, but unchanged at the top of the domain, resulting in a shallower temperature gradient. This difference in temperature gradient will be most noticeable in the continuum centre-to-limb variations, which we discuss below.  

The effect of NLTE in the PHOENIX models seems to be a cooling of the outer layers ($\sim 50$\,K), with minor differences at other depths. This NLTE cooling goes in the opposite direction of the NLTE effects of other PHOENIX solar NLTE models \citep{ShortHauschildt2005,ShortHauschildt2009}, where the NLTE effects cause a warming in the outer layers as naively expected from reduced line blanketing and surface cooling. This discrepancy seems to be associated with a different choice of atomic species treated in NLTE.

\begin{figure*} 
  \centering
  \includegraphics[width=0.49\textwidth]{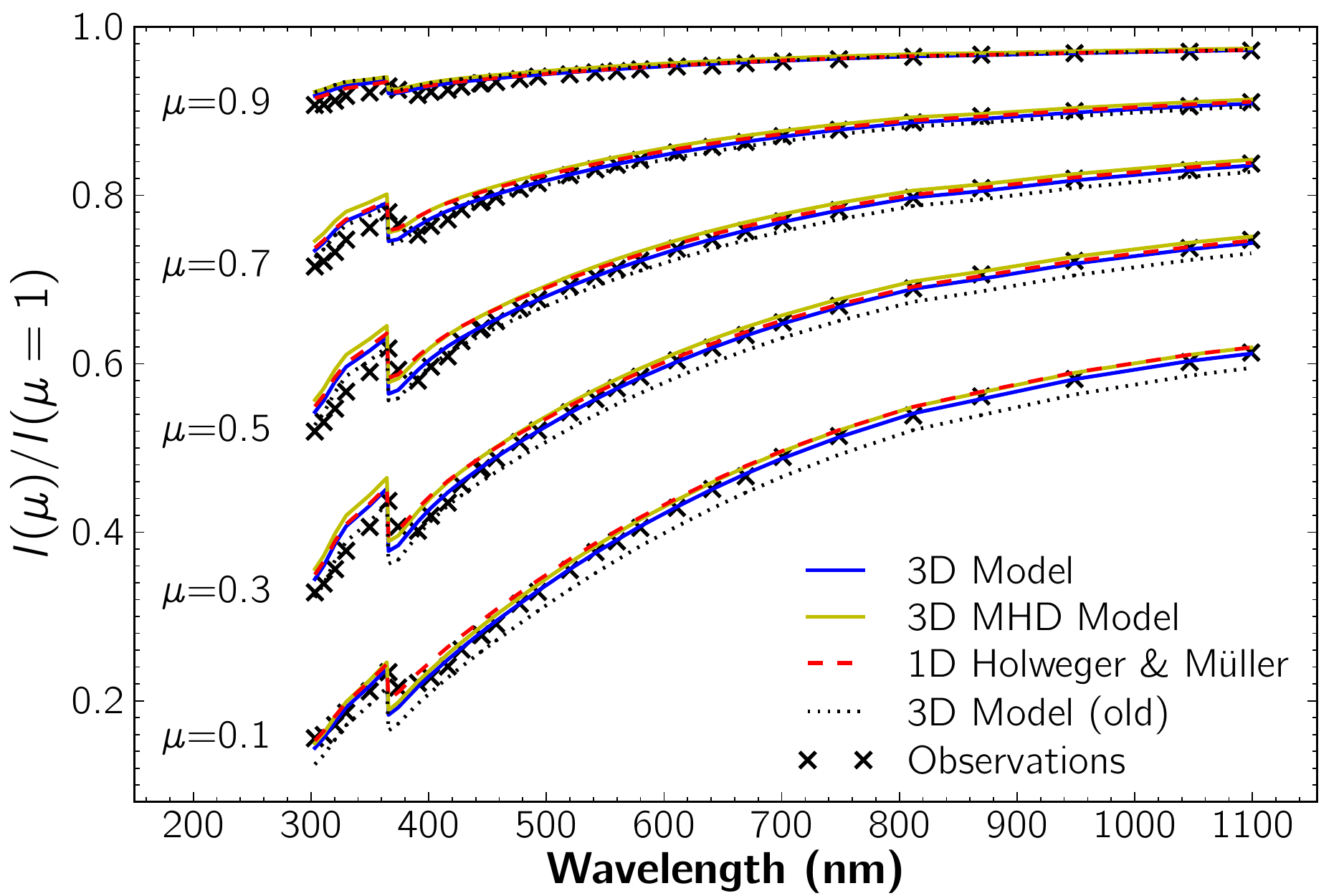}
  \includegraphics[width=0.49\textwidth]{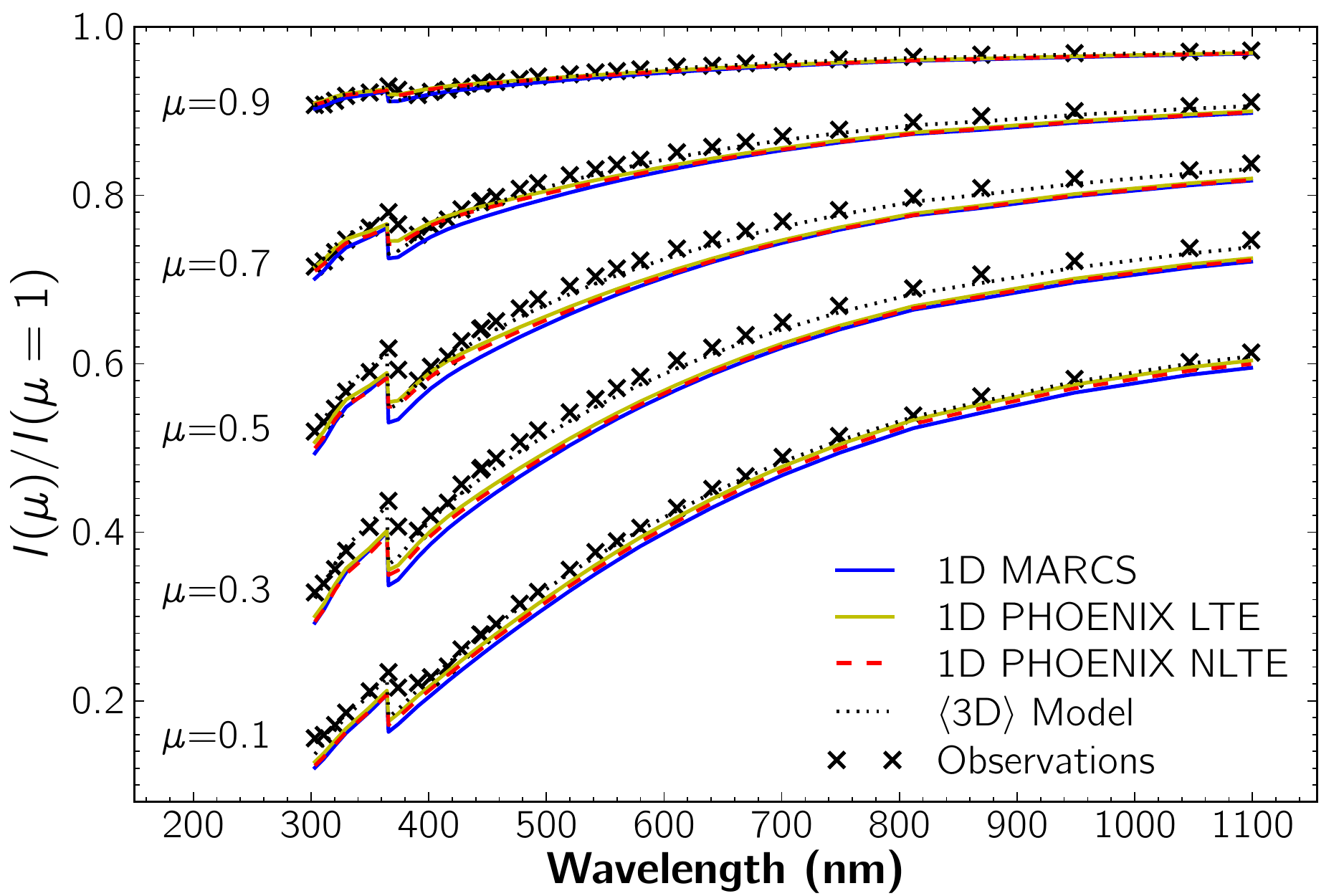}\\
  \includegraphics[width=0.49\textwidth]{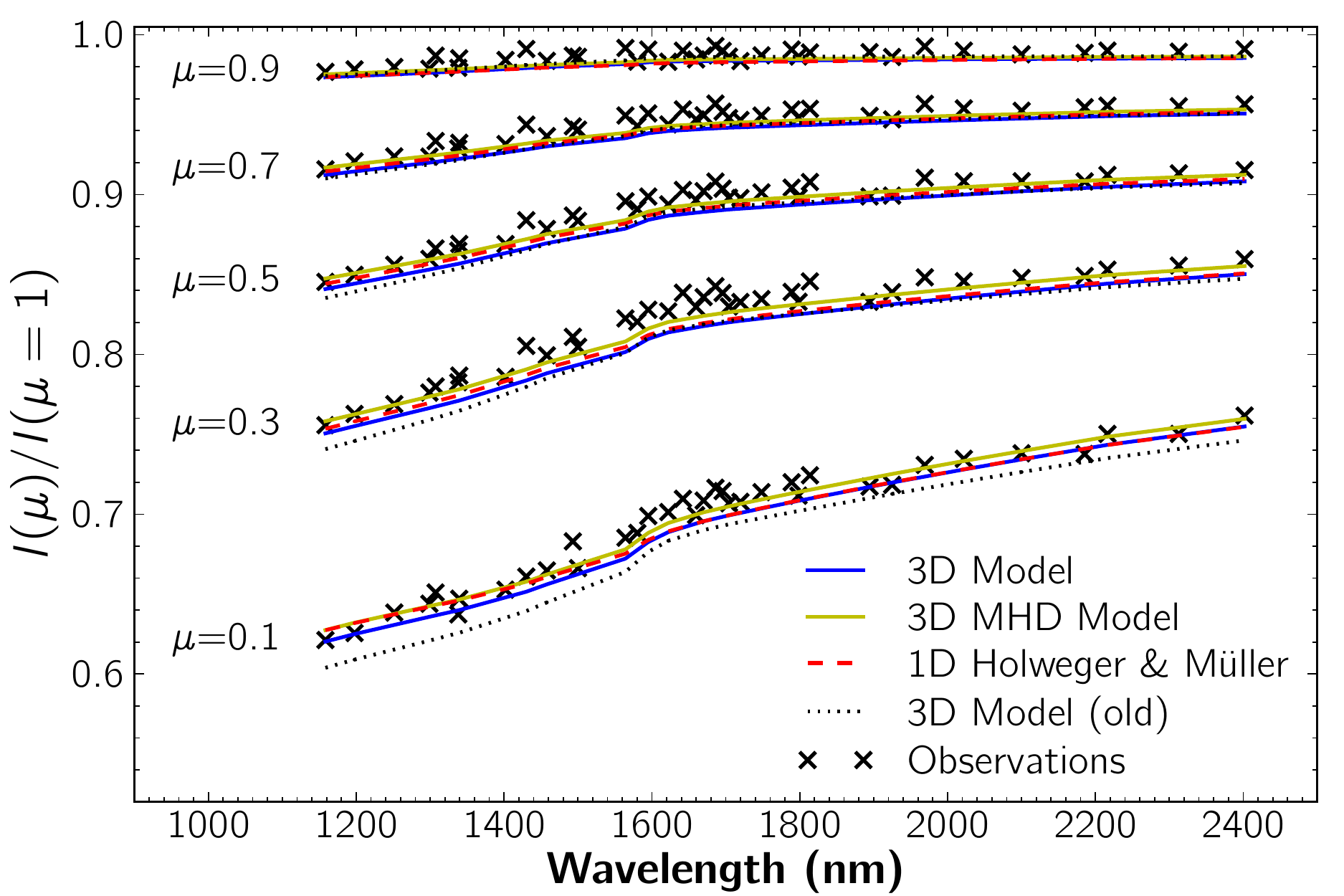}
  \includegraphics[width=0.49\textwidth]{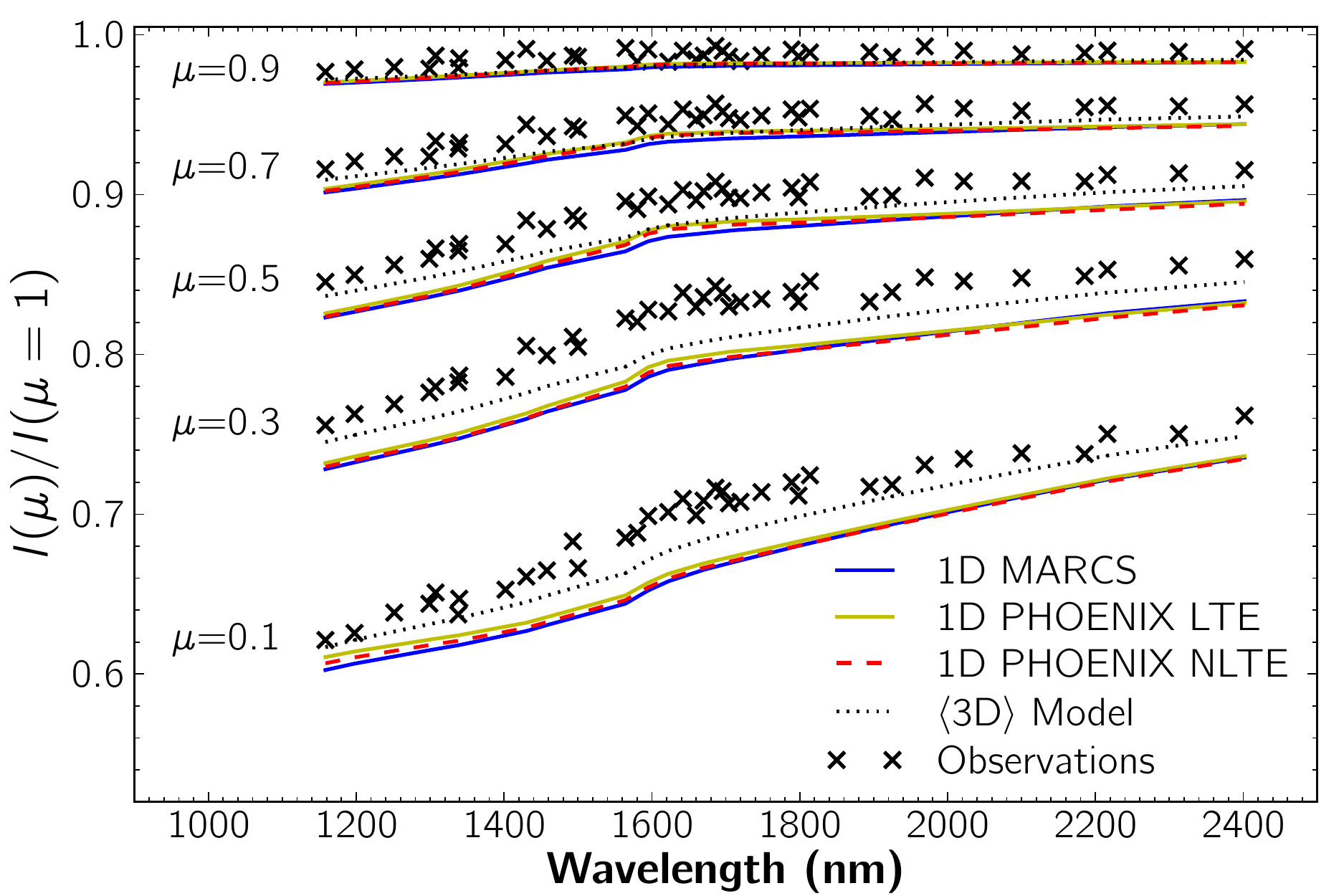}
  \caption{Centre-to-limb variations in the continuum intensity. \emph{Top panels:} comparison with the visible/infrared observations of  \citet{NeckelLabs1994}. \emph{Bottom panels:} comparison with the near-infrared observations of \citet{Pierce1977}, for wavelengths between 1158.35--2401.8~nm.}
  \label{fig:clv_all}
\end{figure*}

When debating on the advantages and disadvantages of employing a full 3D analysis to stellar photospheres, an important question to ask is: is the advantage brought by the 3D treatment of convection merely a realistic temperature stratification, or are the spatial and temporal inhomogeneities essential to derive accurate abundances? To answer these questions we included another 1D model in the testing, corresponding to the spatial and temporal mean structure of the 3D model (shown in Fig.~\ref{fig:ttau}). The horizontal averaging was done on surfaces of equal optical depth rather than geometrical height. This model, hereafter the $\langle$3D$\rangle$ model, will enable us to disentangle the effects of the mean temperature structure from the full 3D treatment.

\section{Continuum centre-to-limb variations\label{sec:clv}}

\subsection{Context}

The centre-to-limb variations (CLV) of continuum intensities provide a sensitive probe of the solar photosphere. Because the continuum intensity is proportional to the local source function of continuum forming regions, its CLV is a measure of the temperature variation with depth (the closer to the solar limb, the higher up in the atmosphere). The variation of depth can be expressed in terms of $\mu\equiv\cos\theta$, where $\theta$ is the heliocentric viewing angle. Normalising $I(\mu)$ by the disk-centre value $I(\mu=1)$, one has a measure of the temperature gradient of the photosphere around the continuum forming layers. This provides a robust test of models.

\subsection{Observations}

We make use of the CLV observations of \citet{NeckelLabs1994} and \citet{Pierce1977}. They cover respectively the wavelengths between \mbox{303--1099~nm} and \mbox{740.4--2401.8~nm}, as shown in Fig.~\ref{fig:clv_obs} for $0.1 \leq\mu\leq 0.9$. We compare the models with \citet{NeckelLabs1994} for $\lambda \leq 1099$~nm and \citet{Pierce1977} for $\lambda > 1099$~nm. Although not used in our comparison, the observations of \citet{Pierce1977a}, covering the range \mbox{303.3--729.7~nm}, are also plotted in Fig.~\ref{fig:clv_obs} and agree very well with \citet{NeckelLabs1994}.

Other CLV observations of longer wavelengths exist, such as \citet{Spickler1996}. However, as is visible in Fig.~3 of \citet{Spickler1996}, there is a considerable scatter between different observations, especially for $\lambda\gtrsim 4\:\:\mu\rm{m}$. Because of these uncertainties we do not include them in this comparison.

\subsection{Results and discussion}

The model predictions were computed using our 3D LTE line formation code, which was used to obtain the predicted intensity at different inclinations. The intensities were computed for nine different values of $\mu\equiv\cos\theta$ and for each inclination except the vertical they were averaged over four $\varphi$-angles in the 3D case. These intensities were interpolated in $\mu$ for the inclinations shown. For the 3D model the intensities were computed for each of the 90 snapshots, the final value being the spatial and temporal average.

\begin{figure} 
  \centering
  \includegraphics[width=0.45\textwidth]{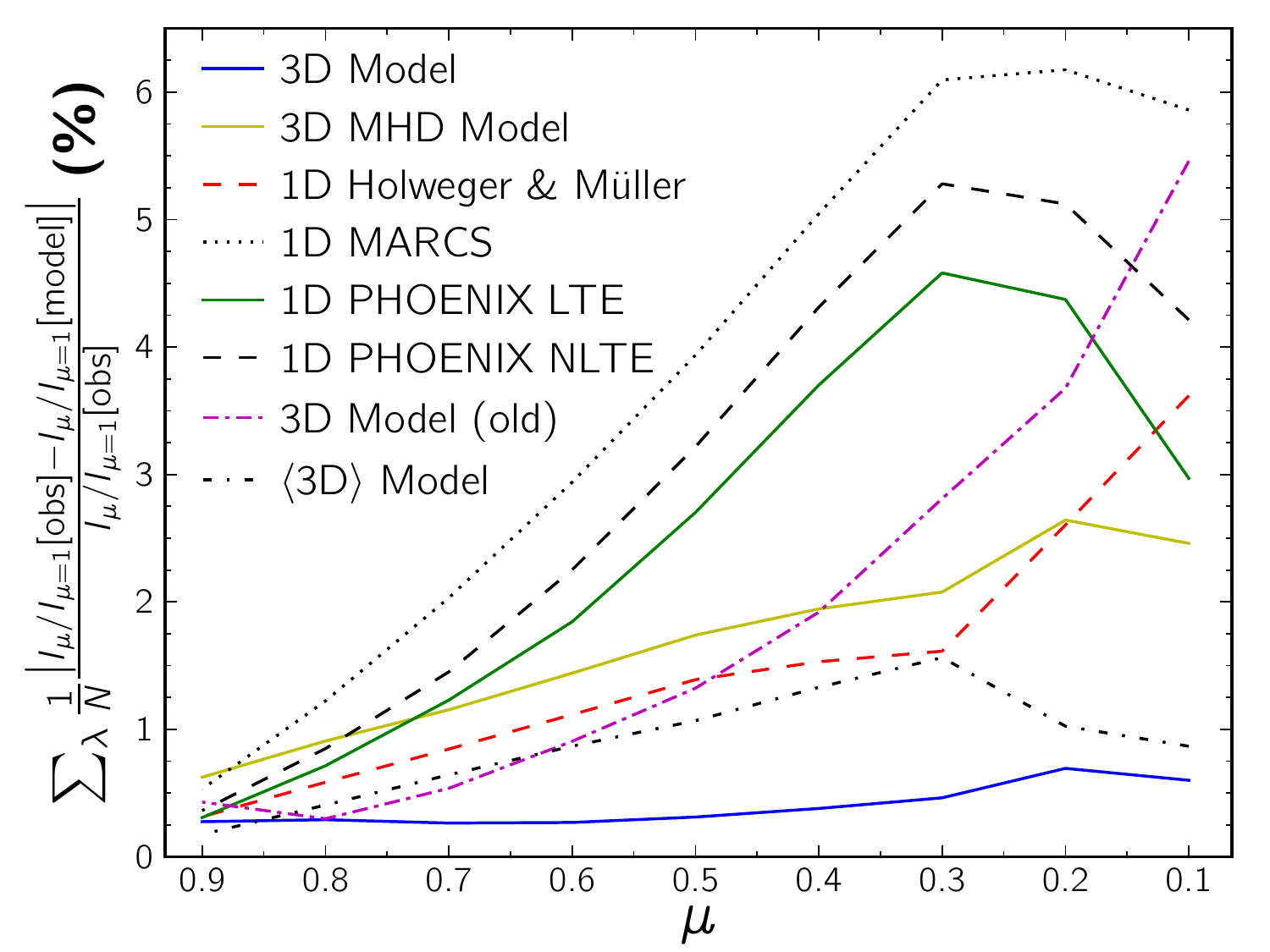}
  \caption{Normalised differences between observations and models in the centre-to-limb variation, averaged over wavelength as a function of $\mu$ (see text). Comparison with \citet{NeckelLabs1994} for \mbox{$400 < \lambda < 1099\:\rm{nm}$}.}
  \label{fig:clv_sum}
\end{figure}

Results for the visible and near-infrared continuum centre-to-limb variation are shown in Fig.~\ref{fig:clv_all}. A more compact comparison with \citet{NeckelLabs1994}  is made in Fig.~\ref{fig:clv_sum}, where we plot the (normalised) absolute value of the difference between models and observations, averaged over wavelength for each $\mu$ value. Because of uncertainty in the observations (from the difficulty in finding continuum regions), this average is limited to the \mbox{$400-1099\:\rm{nm}$} wavelength region.

The results show an excellent agreement between the 3D hydrodynamical model of \citet{Asplund2009} and the observations, particularly when comparing with \citet{NeckelLabs1994}. This agreement is visibly better even than that of the 1D Holweger \& M\"uller model, which is quite remarkable given that it was empirically constructed to fit the centre-to-limb variations. It should be noted, however, that the observational and atomic data have improved much since the construction of the Holweger \& M\"uller model. The application of modern opacities to the Holweger \& M\"uller $T(\tau)$ stratification therefore results in discrepancies not present at its development in 1974.
We note also that the spectral resolving power of the solar atlas employed by  Holweger \& M\"uller was poorer than the atlases available today, leading to less steep line cores and consequently a too high temperature was inferred from the spectral inversion process; rectifying this shortcoming results in a temperature structure very similar to the mean of the here employed 3D model (N. Grevesse, 2009, private communication). With the infrared observations the agreement with the models is slightly worse. However, observations at these wavelengths seem more uncertain, as can be seen by the increased scatter between data points. It is likely that this additional uncertainty affects the agreement with the models. Indicative of this is the region between \mbox{1500--1850~nm}, where the observations show some scatter and are consistently higher than the model predictions.

The inclusion of a 10~mT magnetic field in the 3D model somewhat degrades the agreement with continuum centre-to-limb variation as seen in Fig.~\ref{fig:clv_all} and \ref{fig:clv_sum}. This difference can be traced to the shallower temperature gradient (Fig.~\ref{fig:ttau}) in the MHD model. The 3D MHD model performs slightly worse than the Holweger \& M\"uller model.

The agreement with the theoretical 1D models is not as good. It is interesting to note in Fig.~\ref{fig:clv_sum} that LTE models of MARCS and PHOENIX have the same trend with $\mu$, although the PHOENIX model performs slightly better. The results for the PHOENIX NLTE model depart only slightly from the LTE model results. The NLTE cooling of the outer layers seen in Fig.~\ref{fig:ttau} causes a slightly steeper temperature gradient, which leads to a worse agreement with the observed centre-to-limb variations. The overall structure and dependence with $\mu$ remains essentially the same for both PHOENIX models as well as for the MARCS model, as seen in Figs.~\ref{fig:clv_all} and~\ref{fig:clv_sum}, due to the similarity in $T(\tau)$ for $-2< \log\tau < 0$, the layers largely tested with continuum CLV.

Compared to other models, the differences between the 3D and $\langle$3D$\rangle$ models are small, meaning that the mean temperature gradient is the main driver of the continuum CLV behaviour. Nevertheless, the 3D model predictions agree even closer with the observations, confirming the results of \citet{Koesterke2008}, although we find a smaller \mbox{``3D--$\langle$3D$\rangle$''} difference. Looking at Fig.~\ref{fig:clv_all}, the $\langle$3D$\rangle$ model lies slightly below the 3D model, in other words the effect of the atmospheric inhomogeneities increases \mbox{$I(\mu)/I(\mu=1)$}. One would therefore expect that if spatial and temporal inhomogeneities were added to the Holweger \& M\"uller model, its predictions would lie further above the observations. This indicates that the temperature gradient of the Holweger \& M\"uller model is too shallow compared to the Sun.

We compare also with the old 3D model. While this is a realistic model that reproduces the observed line shifts and shapes \citep{Asplund2000}, its steeper temperature gradient has a noticeable effect on the continuum CLV. Its predictions fare worse when compared to the observations (but still better than the 1D models). We do not use this old model in the other observational tests.

In summary, the confrontation with continuum centre-to-limb observations reveal that the 3D hydrodynamical model is very realistic. The $\langle$3D$\rangle$ hydrodynamical model, the 3D MHD and the Holweger \& M\"uller model all perform slightly worse with too little limb-darkening, while the MARCS and PHOENIX 1D theoretical models all predict much too strong centre-to-limb variation due to a too steep temperature gradient.

\section{Absolute flux distribution\label{sec:flux}}

\subsection{Context}

An independent test of the temperature structure of solar models is
provided by the observed absolute continuum fluxes. They provide an absolute temperature scale for the photospheric layers.

\subsection{Observations}

Several observations of the absolute solar flux/irradiance are available, either space-based \citep[\emph{e.g.}][]{Colina1996,Thuillier2004} or Earth-based \citep[\emph{e.g.}][all obtained with the high-resolution NSO/Kitt Peak Fourier transform spectrometer]{Kurucz1984,BraultNeckelFTS,Kurucz2005}. The space-based observations provide a spectrum clean of terrestrial absorption features, but at the cost of a lower spectral resolution.

\begin{figure*} 
  \centering
  \includegraphics[width=0.45\textwidth]{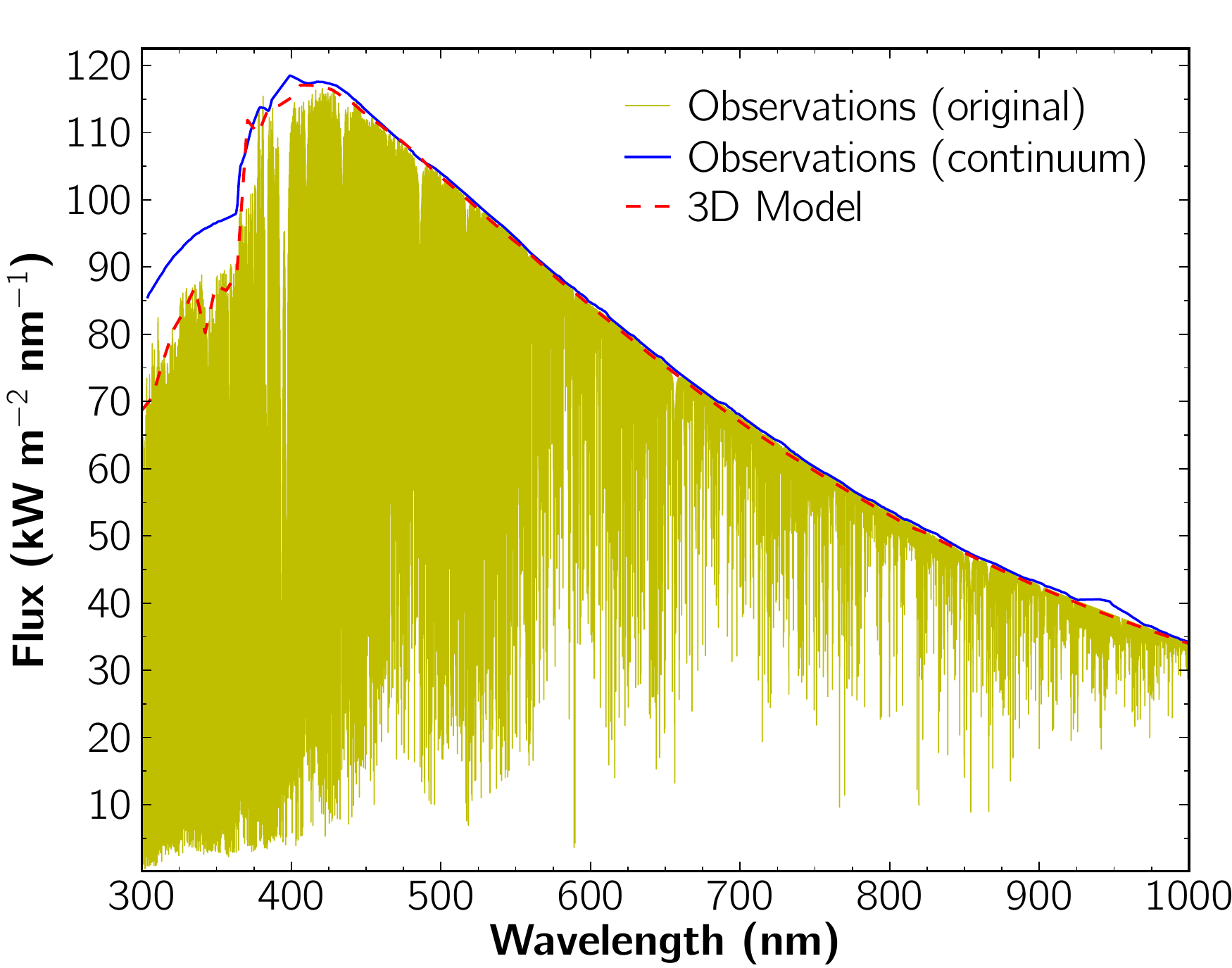} 
  \includegraphics[width=0.45\textwidth]{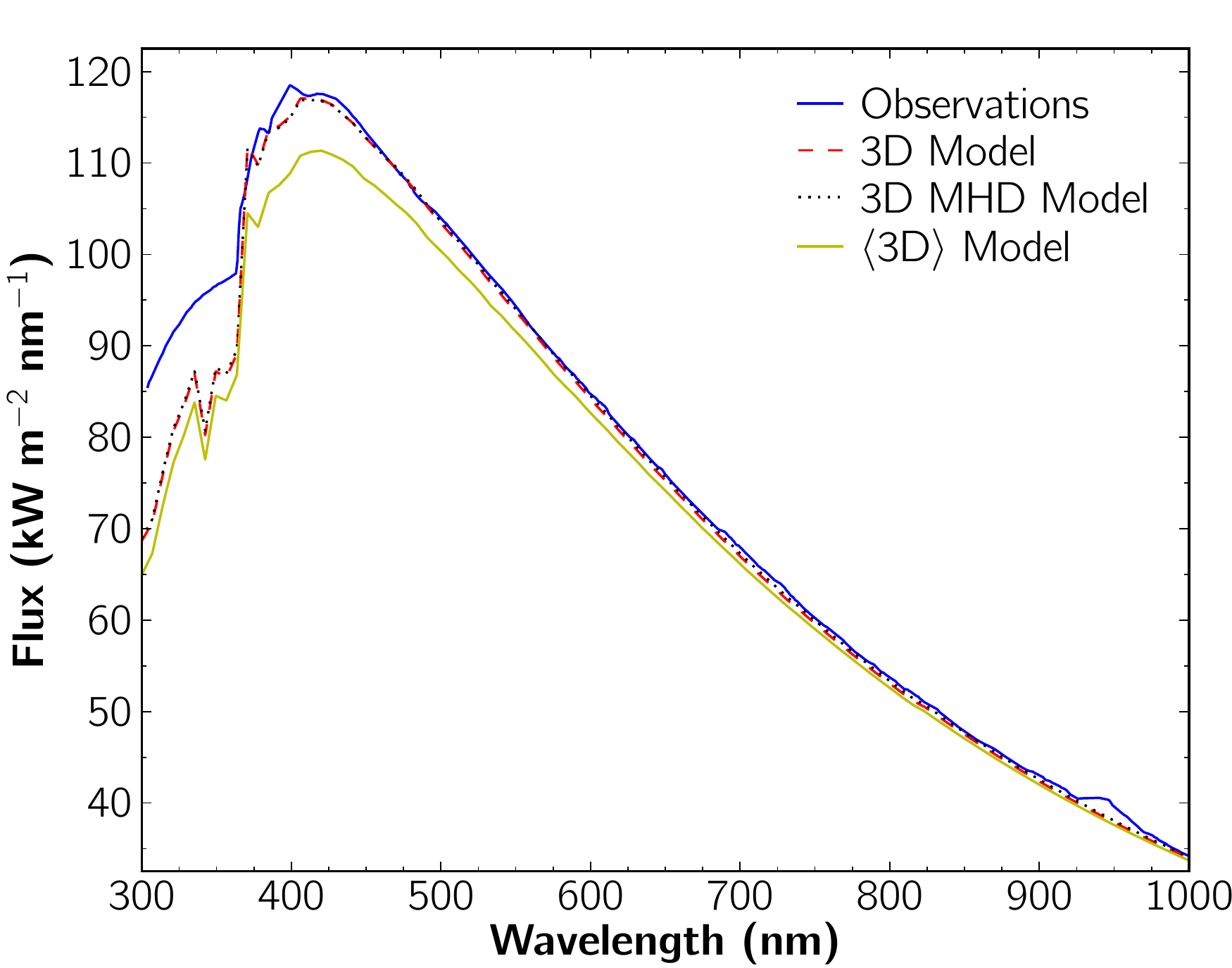} \\
  \includegraphics[width=0.45\textwidth]{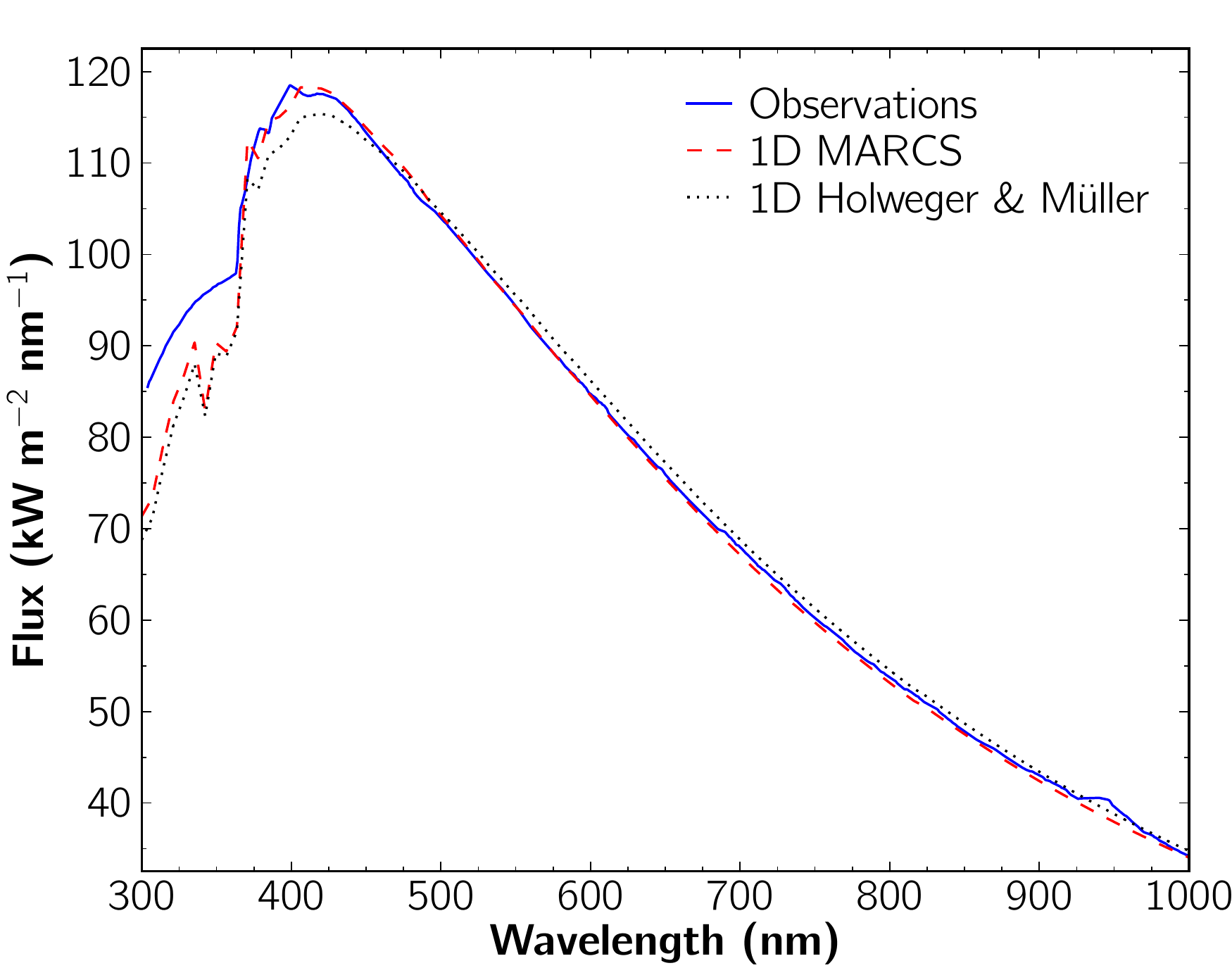} 
  \includegraphics[width=0.45\textwidth]{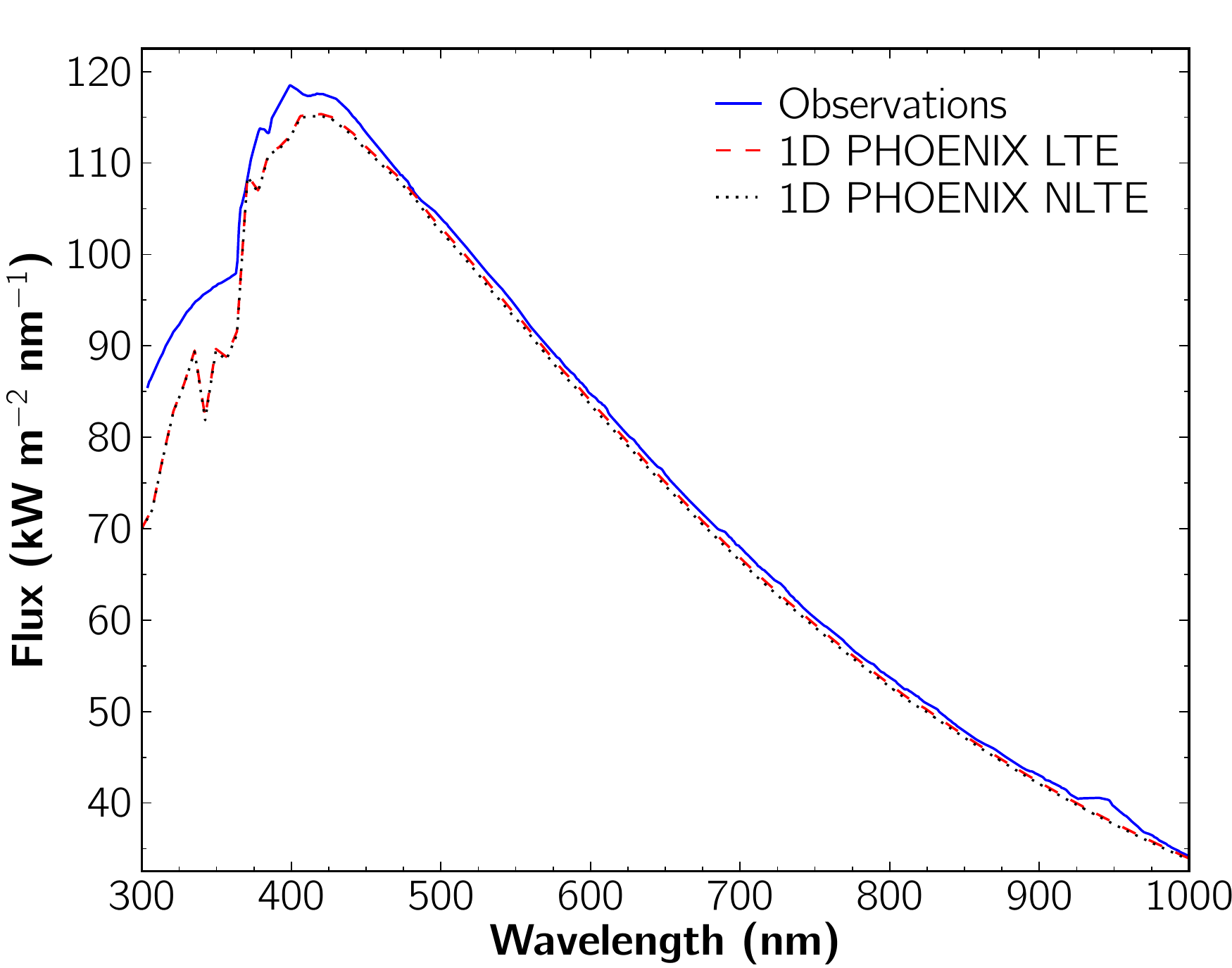} 
  \caption{Absolute fluxes for the models and observations. \emph{Top left:} Comparison of the original fluxes of \citet{Kurucz2005}, our derived continuum fluxes (see text), and the predicted continuum fluxes from the 3D model. \emph{Other panels}: observed continuum fluxes and predicted continuum fluxes for several models.}
  \label{fig:fluxes}
\end{figure*}

To extract the observed continuum fluxes we used an absolute flux atlas divided by a normalised flux atlas (for the points deemed to be near the continuum). We used the \citet{Kurucz2005} irradiances and normalised flux atlases, available for the interval of \mbox{300--1000~nm}. They are a recent reduction of the data used to produce the atlas of \citet{Kurucz1984}, and have been carefully adjusted to remove the telluric absorption features. The choice of using the \citet{Kurucz2005} atlases instead of space-based observations was made because of its consistency between absolute and normalised fluxes.

If another irradiance atlas was used, the slight differences in line strengths or wavelength mismatches between the irradiance and normalised flux atlases would cause some scatter in the continuum fluxes, which would have to be smoothed out \citep[see][where such an approach is followed]{Ayres2006}. To avoid these uncertainties we use the \citet{Kurucz2005} atlases that, being produced from the same data, have a consistent continuum determination and wavelength calibration between the irradiance and normalised flux atlases.

The observed continuum flux was obtained as follows. First the irradiance is converted to flux at the solar surface using the multiplicative factor of $\left[(1\:\: {\rm AU})/R_\odot\right]^2 = 46202$. Then the wavelengths of the (near) continuum points in the normalised flux are identified. These are defined as $F_\lambda > 0.99\; F_\lambda^{\rm{m}}$, where $F_\lambda^{\rm{m}}$ is a local maximum in 5~nm windows. Finally the absolute flux is divided by the normalised flux for the continuum high wavelengths. This ratio is linearly interpolated to a coarser resolution of 1~nm. A few spurious points were manually removed. The \citet{Kurucz2005} irradiance has been normalised to the total solar irradiance of \citet{Thuillier2004}. Following the discussion in \citet{Ayres2006} we have readjusted the continuum fluxes by $-0.4$\%, to account for the more accurate total solar irradiance of \citet{Kopp2005}. In Fig.~\ref{fig:fluxes} we show the original observed fluxes along with our derived continuum fluxes.

\subsection{Results and discussion}

The predicted continuum fluxes were computed with our LTE line formation code. The disk-integrated fluxes are computed using eight $\mu$-angles and four $\varphi$-angles, a total of 32 angles; in addition the disk-centre intensity was calculated. The $\mu$-angles and weights are taken from a Gaussian quadrature and are thus not identical to those used for the centre-to-limb study presented in Sect. \ref{sec:clv}. For the 3D models the fluxes are computed and spatially and temporally averaged for the 90 simulation snapshots. For 1D models we used the same nine $\mu$-angles, including disk centre.

\begin{figure} 
  \centering
  \includegraphics[width=0.45\textwidth]{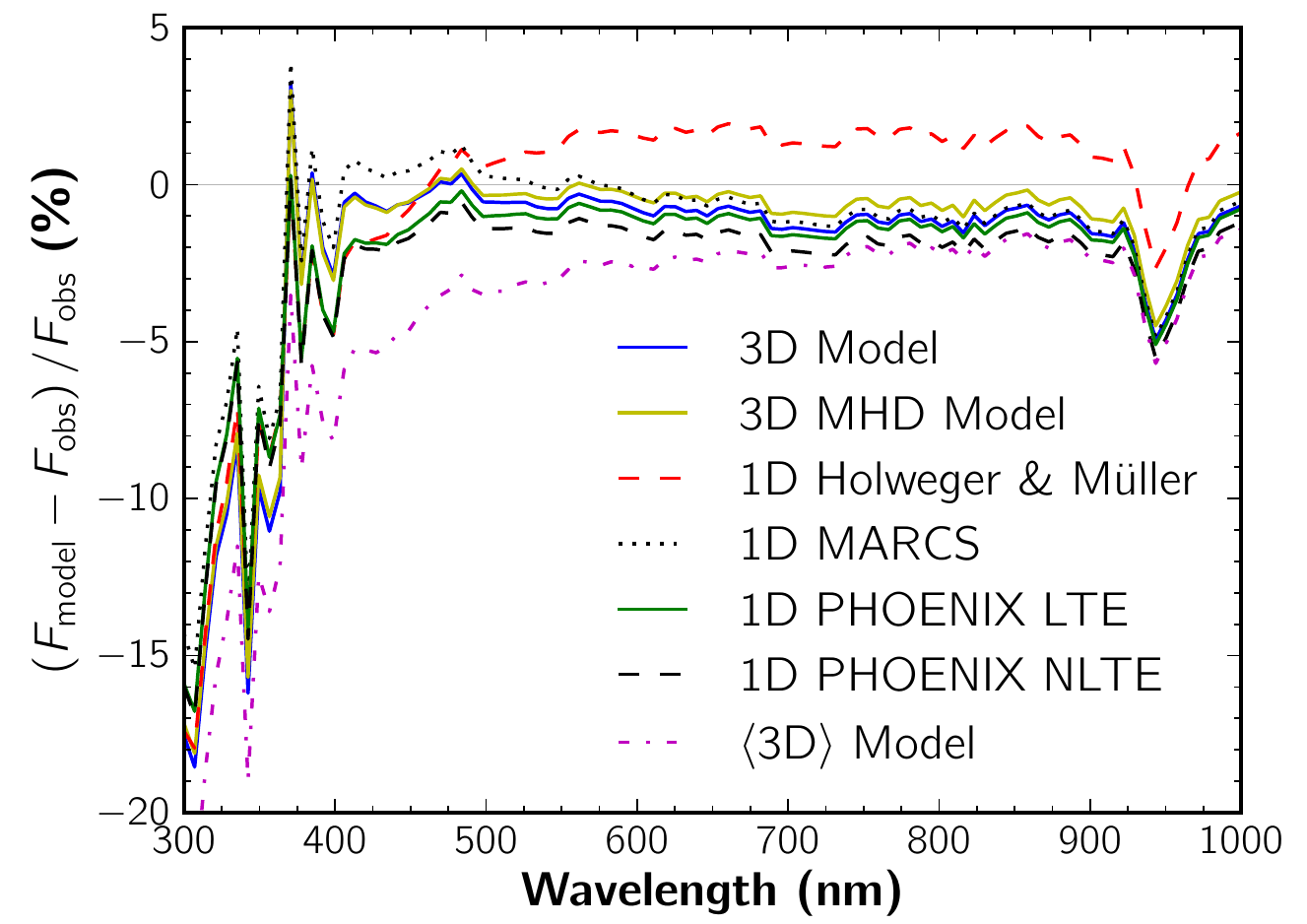} 
  \caption{Continuum flux differences between the models and the observations. For $\lambda \lesssim 450\:\mathrm{nm}$ the observations are not very reliable because of difficulties in continuum placement. The feature at $\lambda\approx 950 \:\mathrm{nm}$ is likely to be caused by uncorrected telluric absorption in the observations.}
  \label{fig:fluxes_diff}
\end{figure}

The resulting fluxes for all models are shown in Fig.~\ref{fig:fluxes}. A differential comparison is also shown in Fig.~\ref{fig:fluxes_diff}. There is an excess of flux for $\lambda\lesssim 370\:\mathrm{nm}$ in the observed continuum fluxes when compared with the original flux (with lines) or any of the models. This difference, also evident in Fig.~\ref{fig:fluxes_diff}, seems to be caused by a too high continuum placement in the normalised \citet{Kurucz2005} flux atlas \citep[\emph{e.g.}, higher than in][]{Kurucz1984,BraultNeckelFTS}. Being a region very crowded with lines, the discrepancy between observations and models likely arises from the difficulty in finding the continuum level (which is systematically overestimated), and not from a failure of the models.

Overall, the 1D MARCS model is the best at reproducing the observed flux, although the differences between different models are small. The Holweger \& M\"uller predictions are consistent with a correct $T_{\rm eff}$, but its flux distribution has a different shape. Between 350--450\,nm it has less flux, but beyond $\lambda\approx 500\,\mathrm{nm}$ it shows an excess flux when compared with the observations. The 3D model consistently predicts slightly less flux than the observations, but nevertheless has a good agreement and reproduces the flux distribution well. The PHOENIX LTE model behaves similarly to the 3D model, but with less flux at the peak of the distribution. The differences between the NLTE and LTE PHOENIX models are very small, not surprising given that very little flux comes from the regions where the NLTE cooling is more efficient.

In the comparison between $\langle$3D$\rangle$ and 3D, one finds considerably less flux coming from the $\langle$3D$\rangle$. This is because the $\tau_{500}$ iso-surface averaging per construction does not preserve the effective temperature. In the 3D model a reasonable amount of flux will be emitted from the snapshots with a higher $T_{\mathrm{eff}}$, as $F\propto T_{\mathrm{eff}}^4$. The $\langle$3D$\rangle$ model is not averaged in $T^4$ and will have a lower $T_{\mathrm{eff}}$. This is perhaps one of the strongest arguments against constructing modified 1D models that recover the mean structure of realistic convection: $T_{\mathrm{eff}}$ will not be preserved. One can argue that the $\langle$3D$\rangle$ model could be averaged in $T^4$. The issue of how to average 3D models will be addressed in a forthcoming paper; tests performed so far indicate that for the purposes of line formation there is little difference between $T$- and $T^4$-averaged models.

The flux differences in Fig.~\ref{fig:fluxes_diff} are quantified in Table~\ref{tab:fluxdiff}, where we show the root mean square differences between the models and observations, summed over the region of 375--975~nm, and normalised to the results from the 3D model. Again one can see that the MARCS model gives the best flux predictions, followed by the 3D MHD and 3D HD model. The PHOENIX and Holweger \& M\"uller models perform similarly while interestingly the $\langle$3D$\rangle$ model shows the largest differences. 

\begin{table}
\caption{Root mean square differences between the models and the observations, between 375--975~nm.}
\label{tab:fluxdiff}
\begin{center}
\begin{tabular}{lr}
\hline\hline
Model & $\Delta{}F^2/\Delta{}F^2_{\mathrm{3D}}$ \\
& \\
\hline
3D Model & 1.00 \\
3D MHD Model & 0.78 \\
1D Holweger \& M\"uller & 3.31 \\
1D MARCS & 0.67 \\
1D PHOENIX LTE & 2.96 \\
1D PHOENIX NLTE & 3.98 \\
$\langle$3D$\rangle$ Model & 15.00 \\
\hline\hline
\end{tabular}
\end{center}
\end{table}

\section{Hydrogen lines\label{sec:hlines}}
\subsection{Context}

The first two lines in the hydrogen Balmer series lines, \ha\ and \hb{}, have traditionally been used for temperature determination in late-type stellar atmospheres. Their wings are typically formed in the region around $-2\lesssim \log\tau_{500}\lesssim 0.5$, deeper than most of the other spectral lines in late-type stars \citep{Fuhrmann1993}, while their cores are formed in the chromosphere. %
The fact that for late-type stars these lines are more sensitive to temperature than they are to gravity, metallicity or indeed the hydrogen abundance has established them as a popular tool for temperature determinations \citep{Fuhrmann1993,Fuhrmann1994,Barklem2002,Behara2009}. The shapes of hydrogen lines, because of their large depths of formation, are however dependent on the convection efficiency \citep[e.g.][]{Ludwig:2009}.

Also relevant to this work are the lines of the Paschen series. With $E_{\mathrm{low}}=12.088\:\mathrm{eV}$, these lines are formed even deeper than the Balmer series. However, owing to their longer wavelengths and lower level populations (and consequently weaker wings), they have not been used as extensively as \ha\ and \hb. We include in our comparisons the Pa$\gamma$ and Pa$\beta$ lines, which are just inside the wavelength range of the high-resolution FTS atlases of the Sun. 

Departures from LTE can be important for hydrogen lines  \citep{Przybilla2004,Barklem2007HNLTE}. While the far wings are formed in deep, hot regions where LTE conditions largely prevail, the LTE assumption is no longer valid for the cores of the strong lines. However, \citet[][hereafter \citetalias{Barklem2007HNLTE}]{Barklem2007HNLTE} shows that in the case of Balmer lines, the NLTE effects can even extend into the line wings, depending on the adopted rates for collisions with H atoms. In their LTE and NLTE study of solar H lines, \citet{Przybilla2004} find a weakening of the NLTE effects for the Paschen lines. To obtain realistic line profiles, we perform NLTE line synthesis as detailed in Sect.~\ref{sec:hsynthesis}.

\subsection{Observations}

For the solar observations of hydrogen lines we use the high-resolution FTS normalised flux atlas of \citet{Kurucz2005}, and the FTS disk-centre atlas of \citet{BraultNeckelFTS}. For our comparison we study the Balmer lines \ha\ and \hb\, along with the Paschen lines Pa$\gamma$ and Pa$\beta$. 

In addition to the disk-centre atlas, \citet{BraultNeckelFTS} have also made available a flux atlas. For the lines considered here it is essentially identical to the \citet{Kurucz2005} atlas, which is not surprising since it was produced from the same raw FTS data. However, we would like to point out that the flux atlas of \citet{BraultNeckelFTS} has a slightly different normalisation, with a lower continuum level. This difference amounts to $\approx 0.4-0.6\%$ of the normalised flux. For a consistent comparison of the flux and disk-centre intensity profiles, we have increased the continuum level of the \citet{BraultNeckelFTS} disk-centre atlas by 0.5\%, so that it matches that of the Kurucz flux atlas. This choice of continuum will not affect the differential results between models, only their relative standing to the observations. Our choice of continuum falls on the Kurucz atlas because it is a more recent (and hopefully more accurate) reduction of the same data. However, the 0.5\% continuum error, which corresponds to a $T_{\mathrm{eff}}$ difference of $\approx 40$~K in the Balmer lines, is within the uncertainties of this analysis (including the errors from the  observations, model, broadening, NLTE effects, etc.). \citet[][hereafter \citetalias{Barklem2002}]{Barklem2002} discuss in detail the several uncertainties associated with the calculation of hydrogen lines, to which one should also add the uncertainties from the choice of hydrogen collisions \citepalias{Barklem2007HNLTE}.

The Pa$\beta$ line, at 1281.8 nm, is not available in the \citet{BraultNeckelFTS} disk-centre atlas nor in the \citet{Kurucz2005} atlas, which only extend to 1250.8~nm and 1098~nm, respectively. For this line we use only the \citet{Kurucz1984} flux atlas, which is based on the same raw data as \citet{Kurucz2005}, but covers a somewhat larger wavelength range. However, the normalisation of the \citet{Kurucz1984} atlas in the region around Pa$\beta$ is problematic. To compensate for this we re-normalised this part of the atlas, finding a suitable continuum level from a polynomial fit to nearby continuum high points.

\subsection{Synthetic profiles\label{sec:hsynthesis}}

\begin{figure} 
  \centering
  \includegraphics[width=0.48\textwidth]{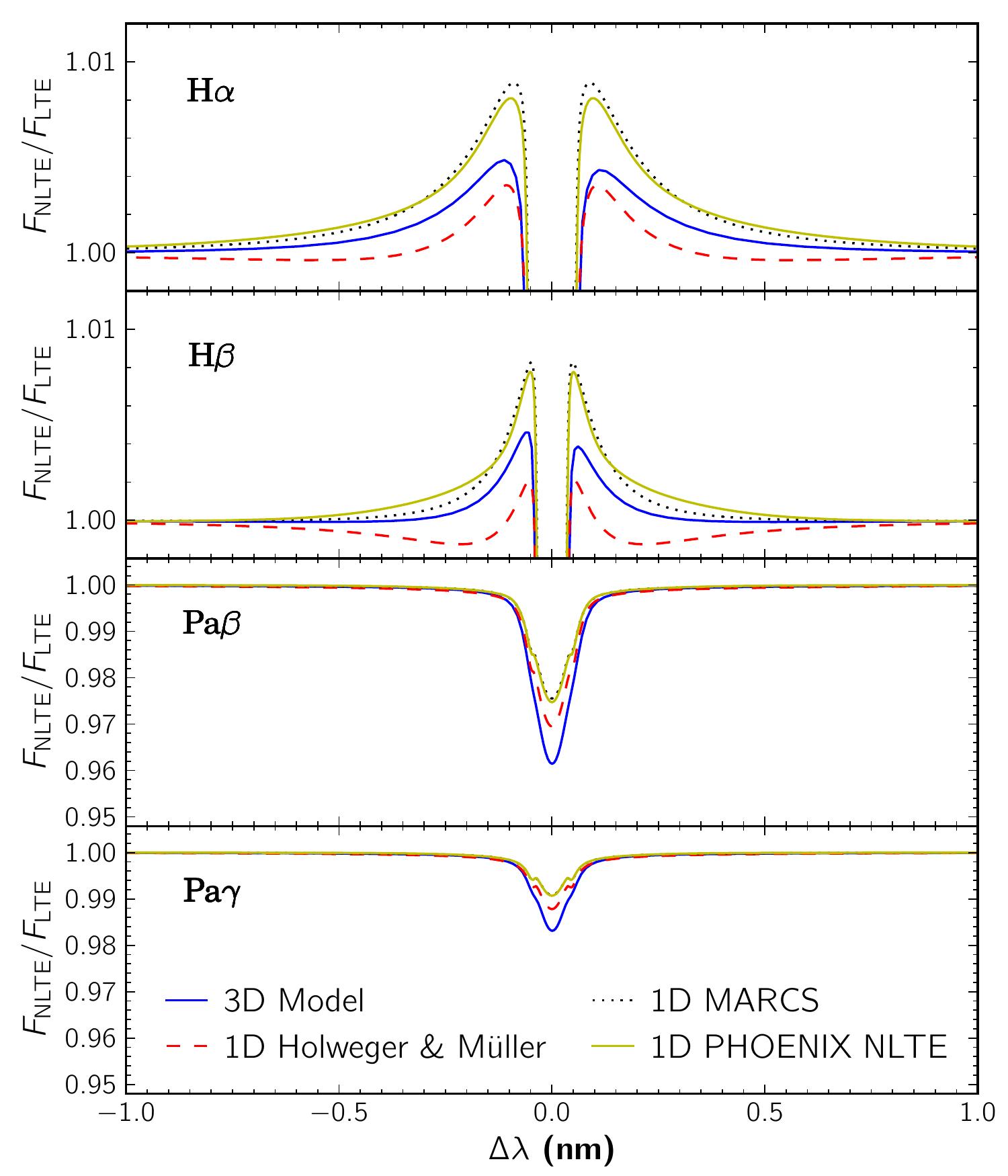} 
  \caption{Flux ratios between NLTE and LTE line profiles, for the hydrogen lines considered and for the different models. Wavelength difference $\Delta\lambda$ measured from the line core. Results for the PHOENIX NLTE model atmosphere (not shown) are indistinguishable from the corresponding LTE model in this figure. For \ha\ and \hb\ the ratio for the line core is not shown, to emphasise the NLTE effects in the wings, formed in the photosphere.}
  \label{fig:hlines_ratios}
\end{figure}

The computation of the H line profiles has been done allowing for departures from LTE. We use a 20-level hydrogen model atom (19 H\,\textsc{i} levels plus continuum) based on the atom of \citetalias{Barklem2007HNLTE}, which was adapted from \citet{CarlssonRutten1992}. The collisional cross-sections by \citetalias{Barklem2007HNLTE} are employed. These include inelastic collisions with electrons and hydrogen atoms \citep[using the cross sections of][assuming \mbox{$E=E_{\mathrm{cm}}$}]{Soon1992}, mutual neutralisation and Penning ionisation. To speed up the calculations, especially in the 3D case, we have neglected the bound-bound transitions starting with a lower level $n>=6$, thus including only 80 bound-bound transitions (out of a total of 171 in the original model atom). Tests with 1D models show that the effects of removing these lines are negligible. 

Line profiles were computed using our LTE code and the MPI-version of the MULTI3D code \citep{Multi3d2009}. To save computational time, full 3D NLTE line formation was performed only on eight snapshots of the simulation. Using MULTI3D to compute the LTE and NLTE line profiles we obtain the wavelength-dependent NLTE/LTE ratio, averaged for these eight snapshots. The final line profiles are then obtained by using our LTE code to compute the 3D LTE line profiles for all the 90 snapshots in the simulation and then multiplying them by the average NLTE/LTE ratio for each wavelength. The very small variation between the NLTE/LTE ratios for the eight snapshots indicates that this procedure is a very good approximation. For the 1D models the computational requirements are unimportant, and they are not time-dependent, but for consistency we use the same procedure of the NLTE/LTE ratio for each model.

\begin{figure*} 
  \centering
  \includegraphics[width=0.49\textwidth]{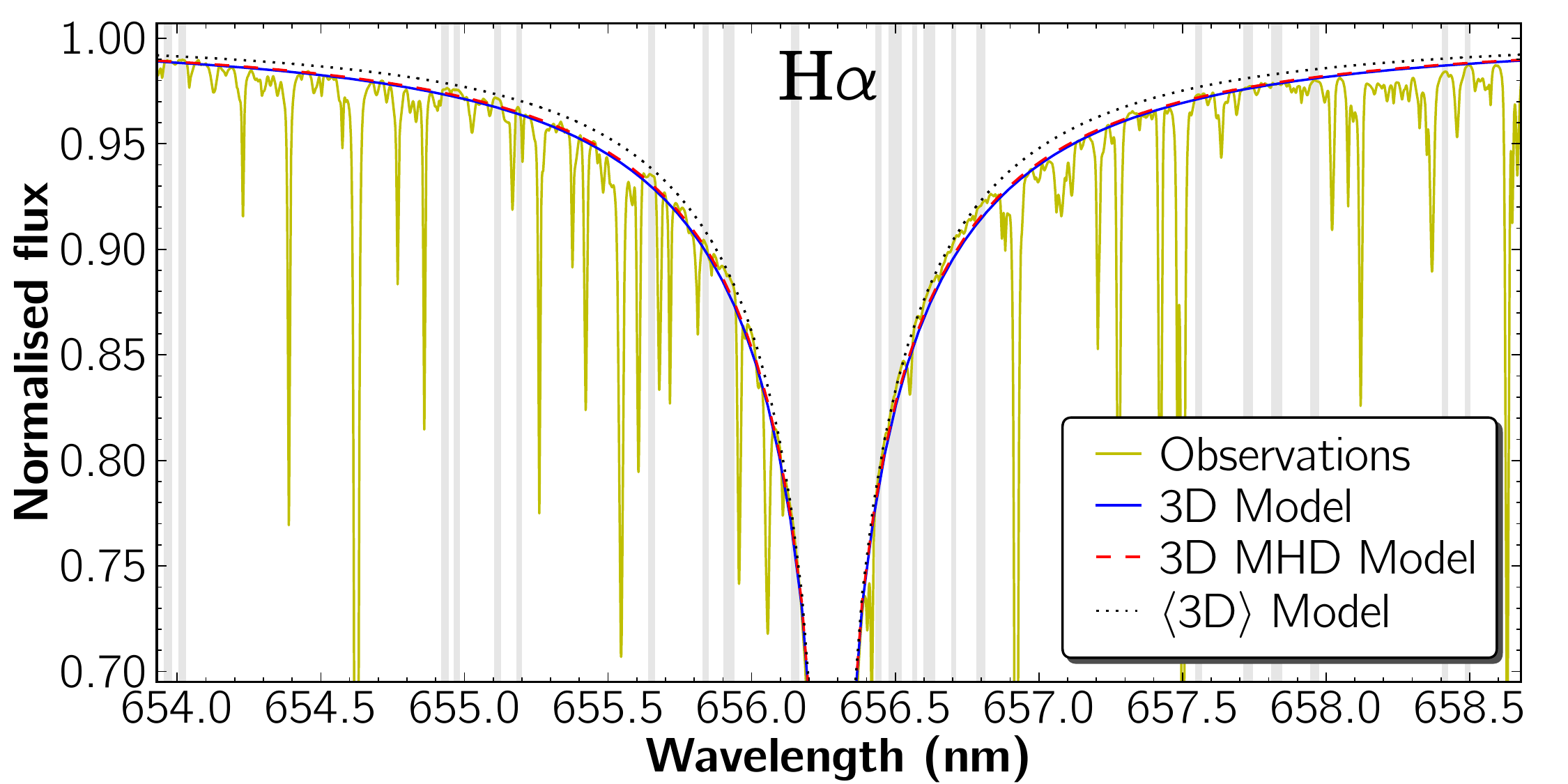} \includegraphics[width=0.49\textwidth]{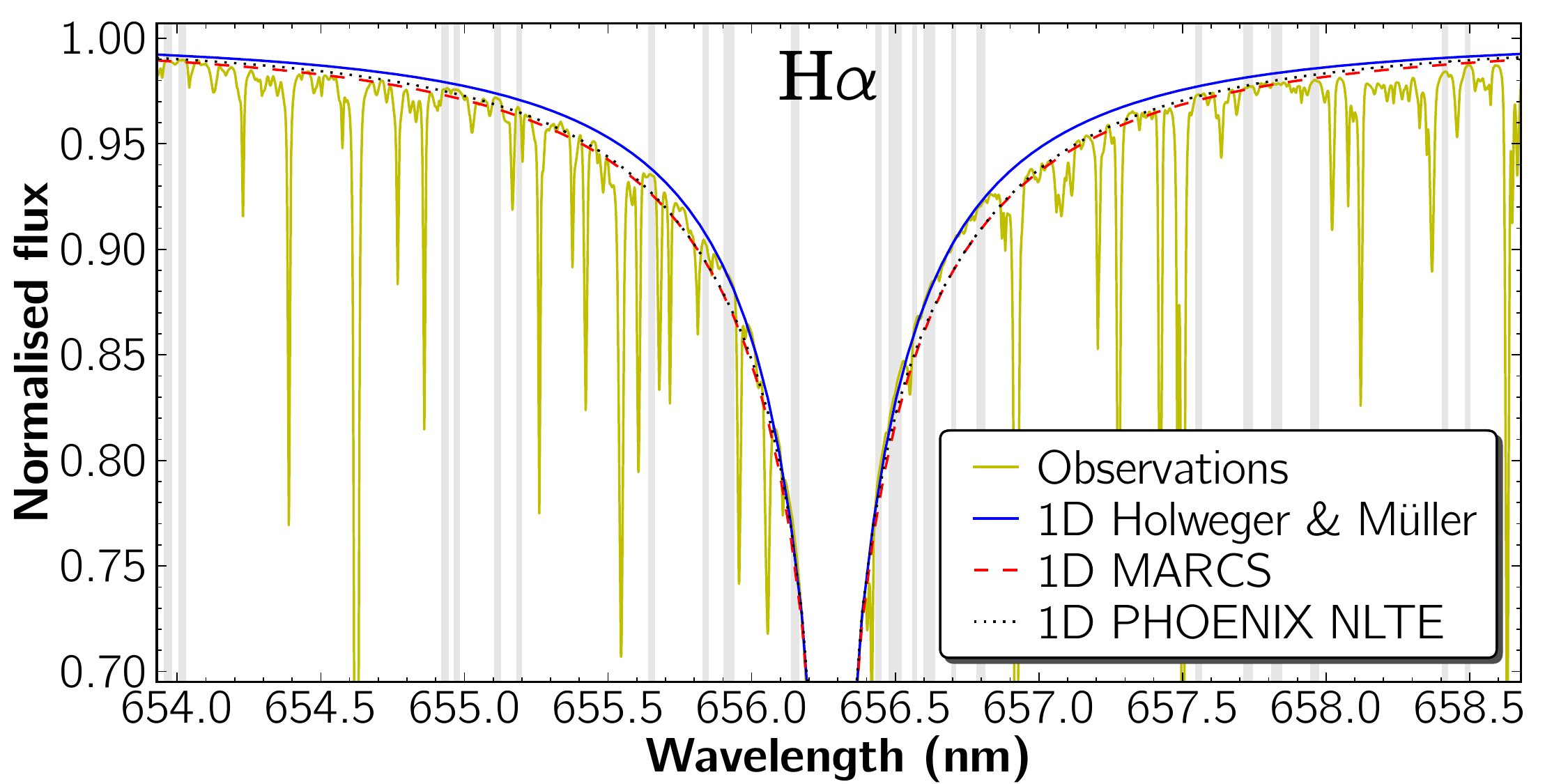}\\
  \includegraphics[width=0.49\textwidth]{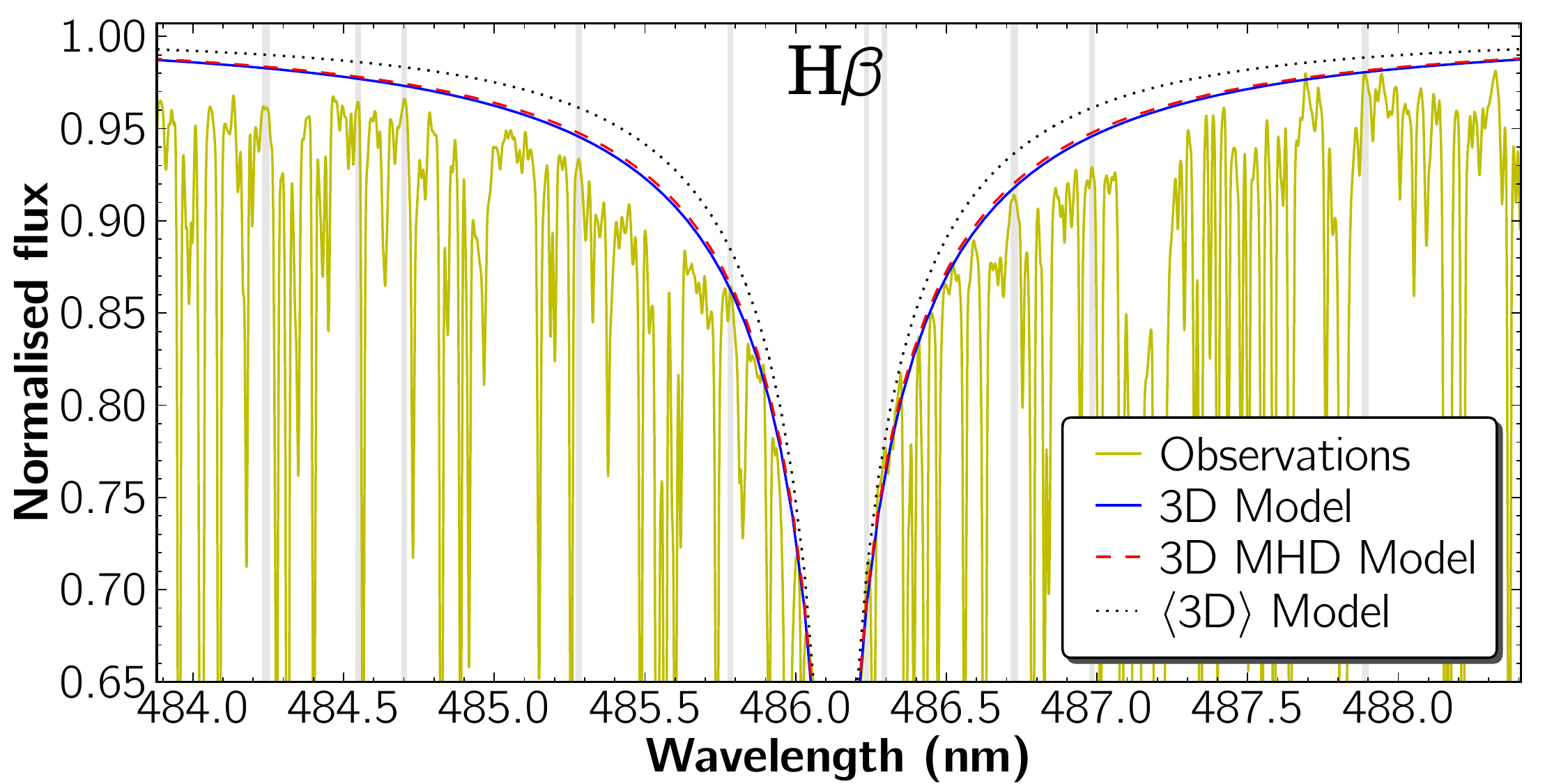} \includegraphics[width=0.49\textwidth]{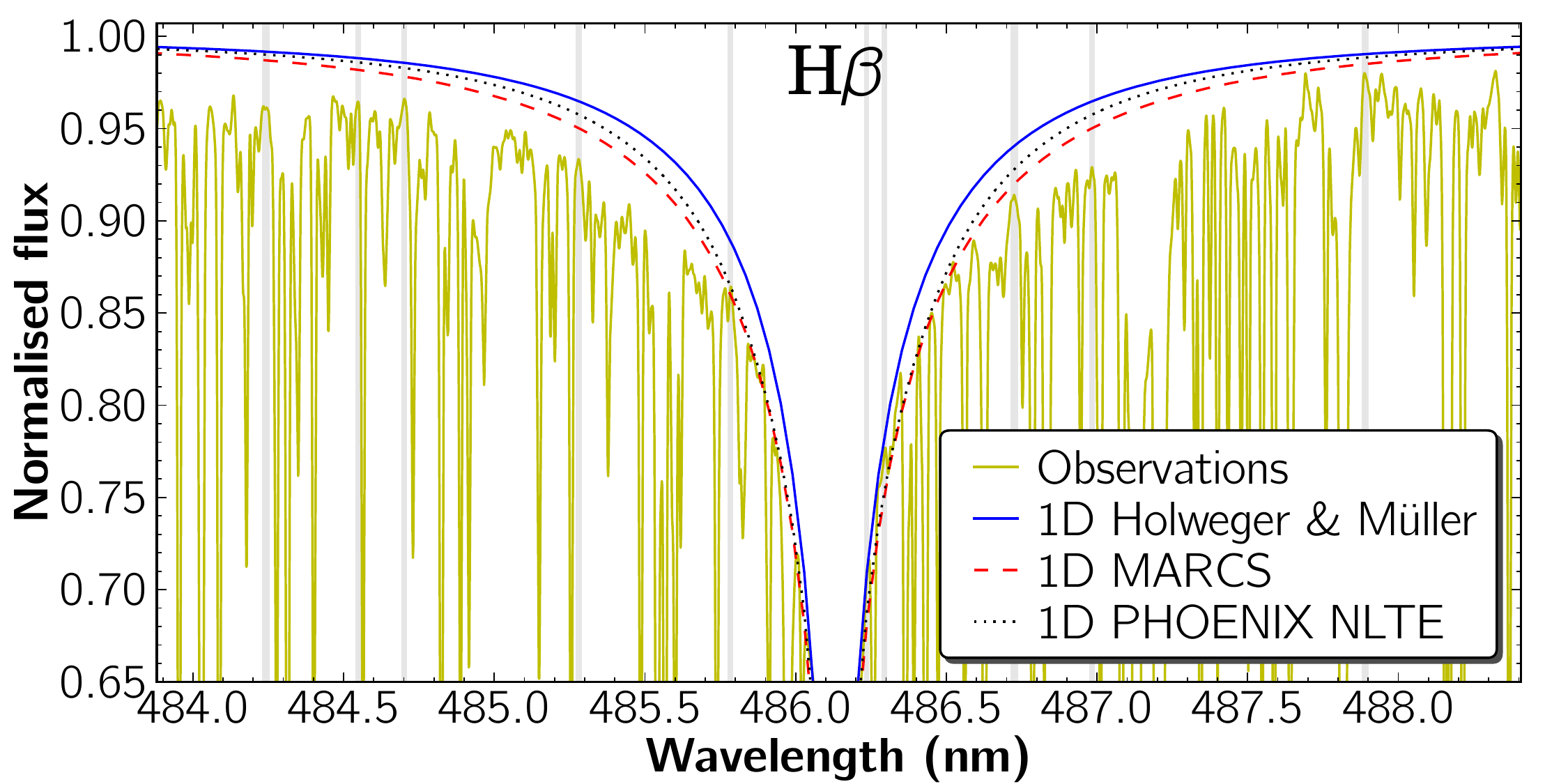}\\
  \includegraphics[width=0.49\textwidth]{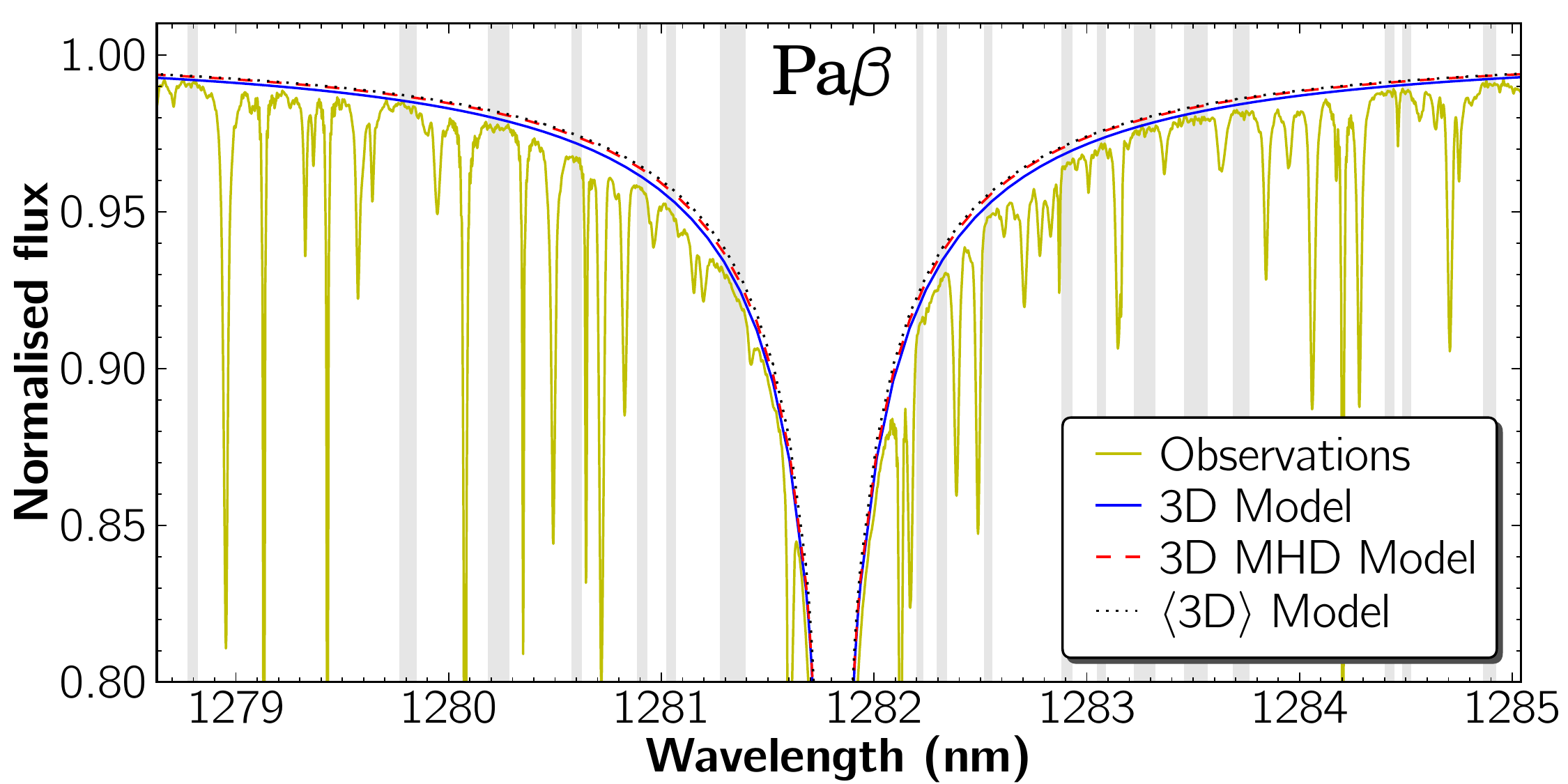} \includegraphics[width=0.49\textwidth]{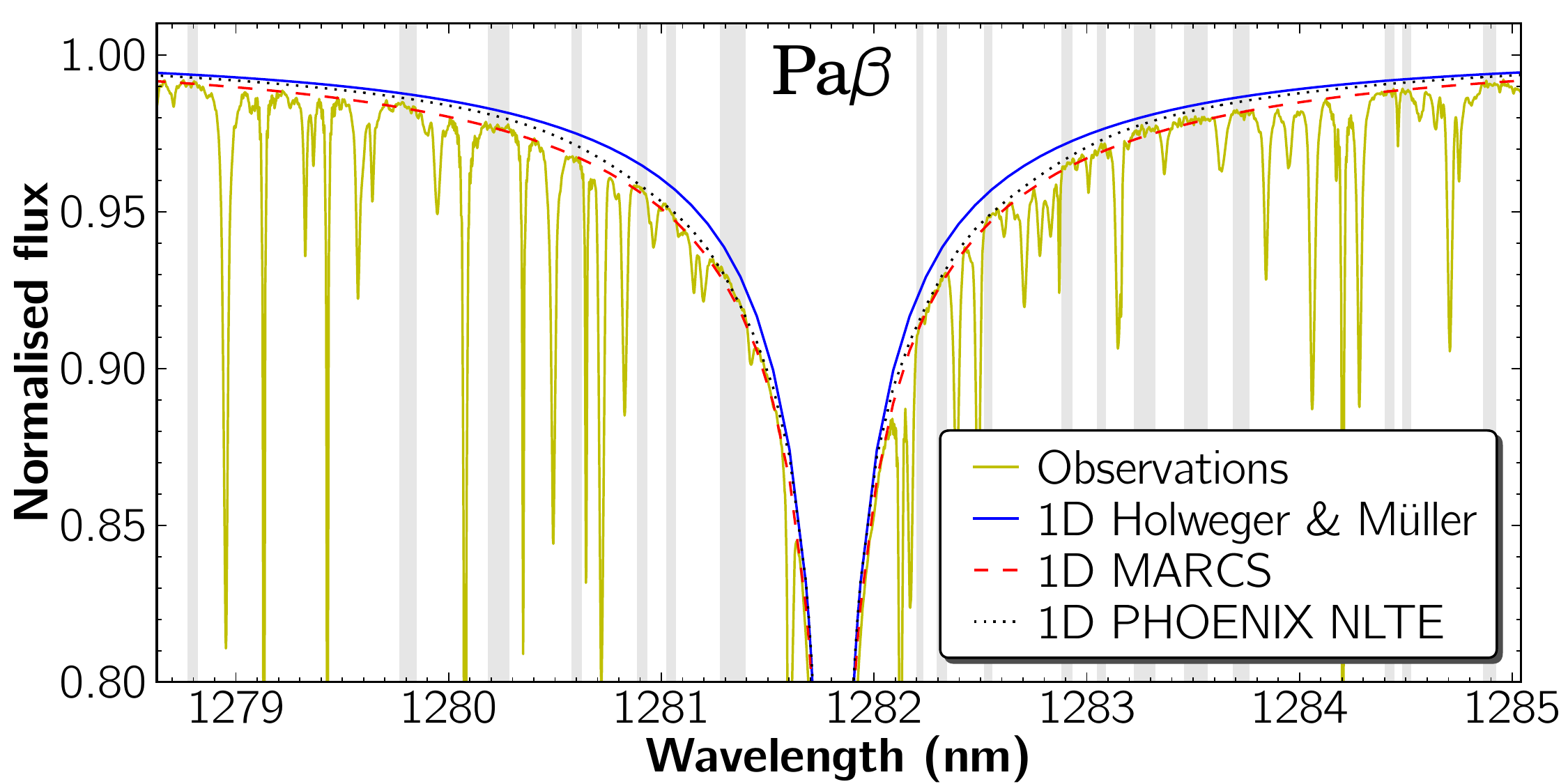}
  \includegraphics[width=0.49\textwidth]{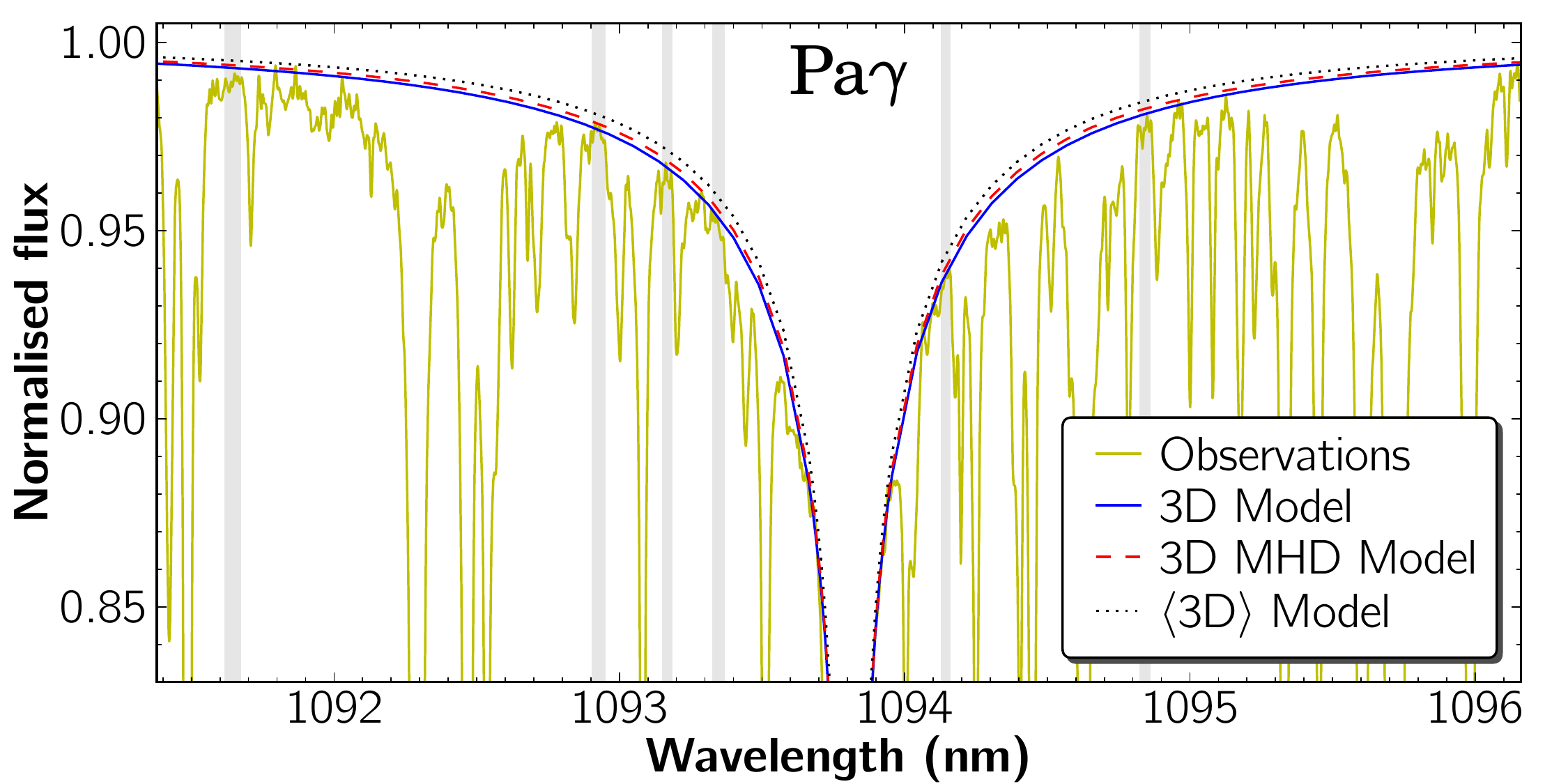} \includegraphics[width=0.49\textwidth]{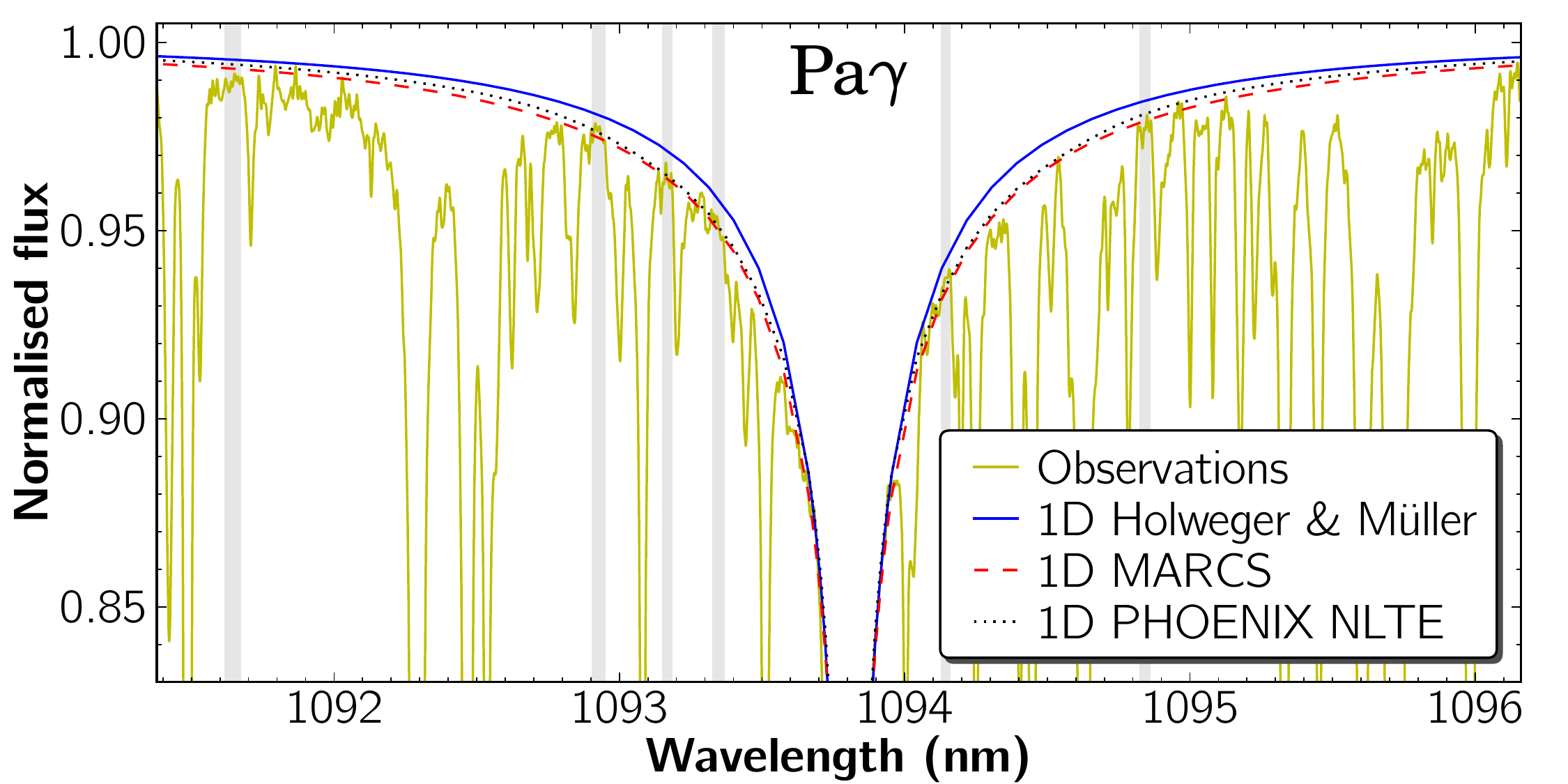}\\
  \caption{Normalised flux profiles for the H lines \hb, \ha, Pa$\gamma$, and Pa$\beta$. Compared with the solar observations of \citet{Kurucz2005}. The exception is Pa$\beta$,  where the \citet{Kurucz1984} atlas is used because the \citet{Kurucz2005} atlas does not cover these wavelengths. For this line the continuum was re-normalised (see text). Synthetic profiles were computed in NLTE. The regions used in the $\chi^2$ analysis are indicated in grey.}
  \label{fig:Hlines_flx} 
\end{figure*}

\begin{figure*} 
  \centering
  \includegraphics[width=0.49\textwidth]{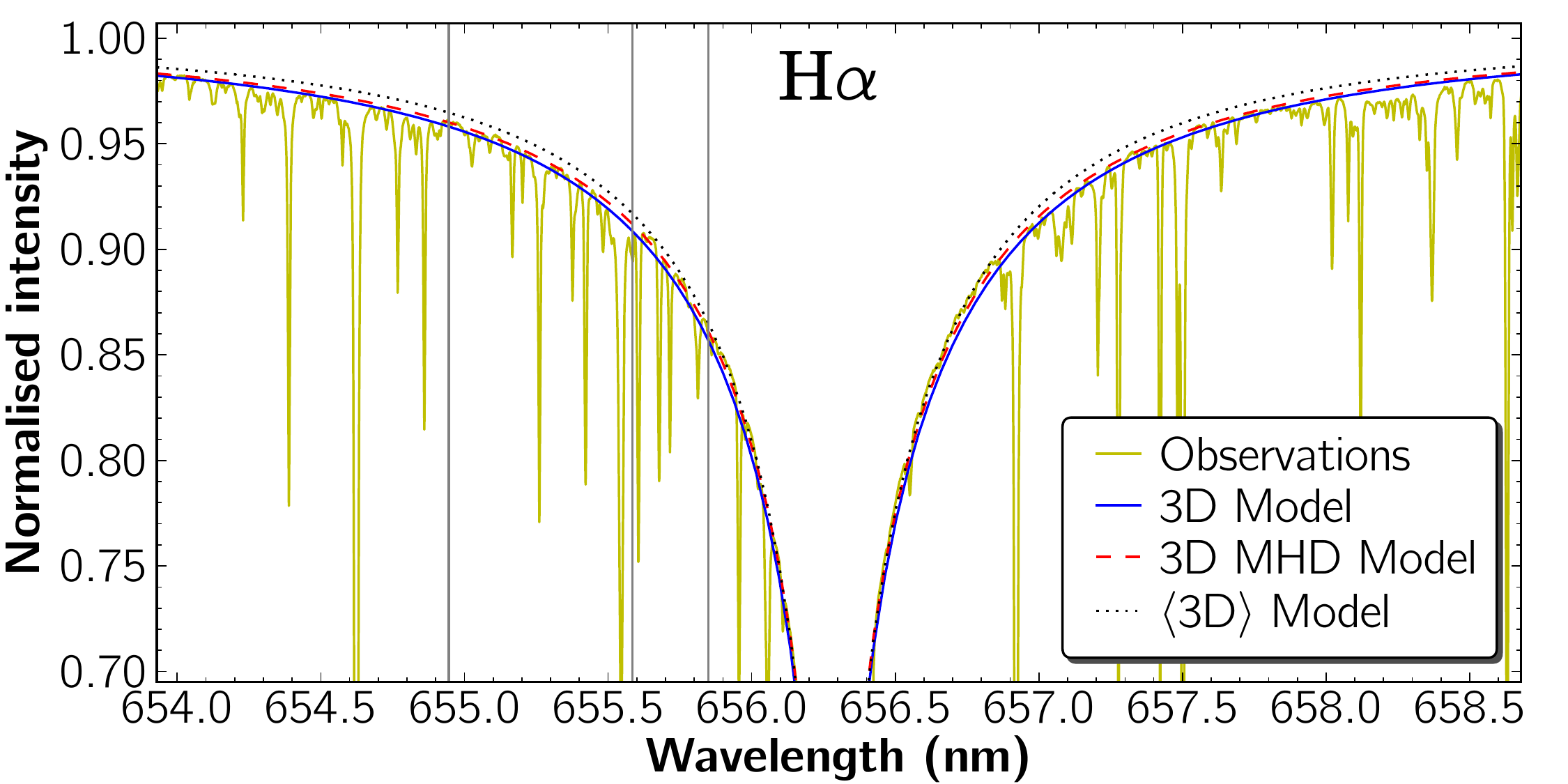} \includegraphics[width=0.49\textwidth]{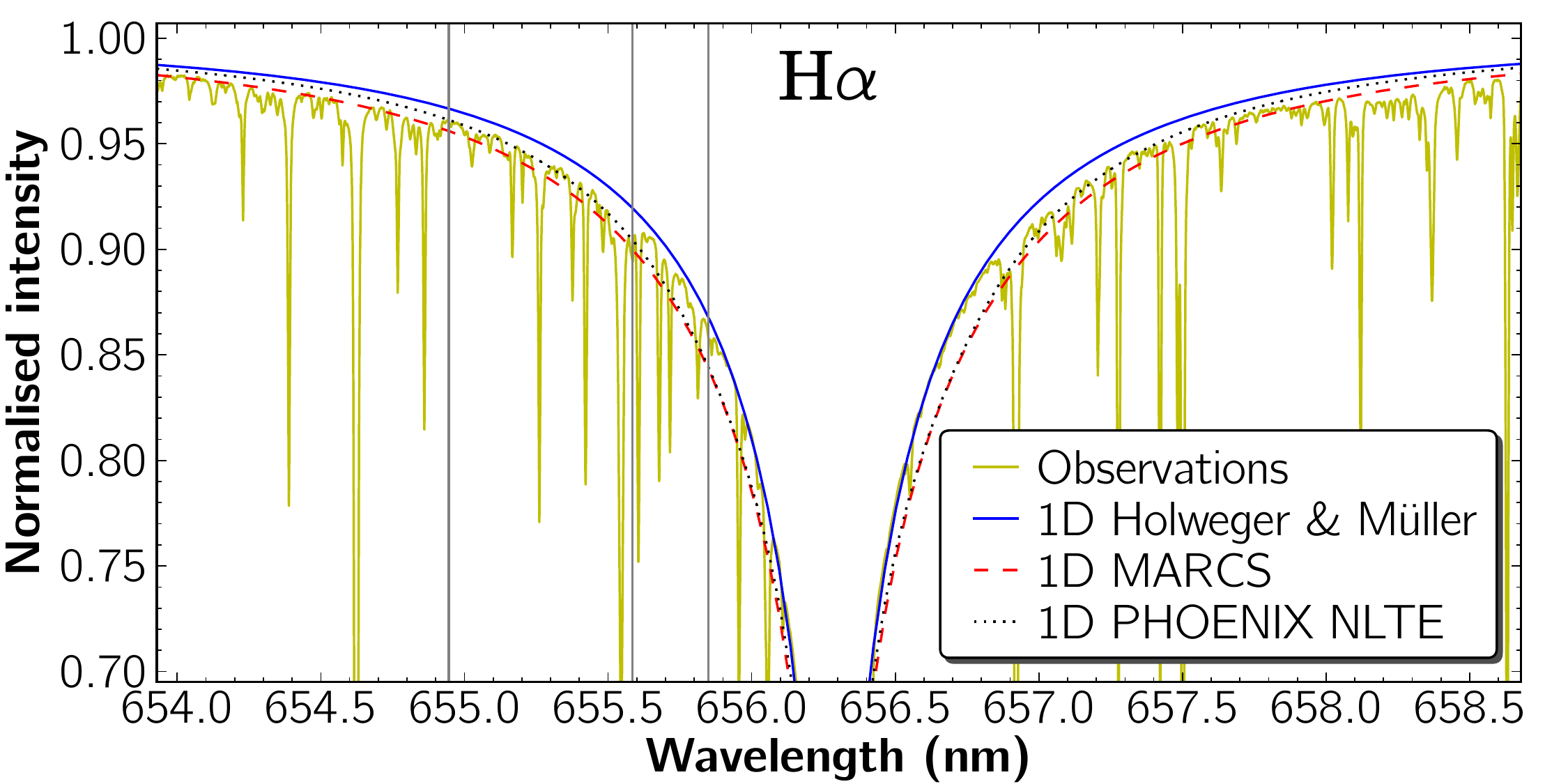}\\
  \includegraphics[width=0.49\textwidth]{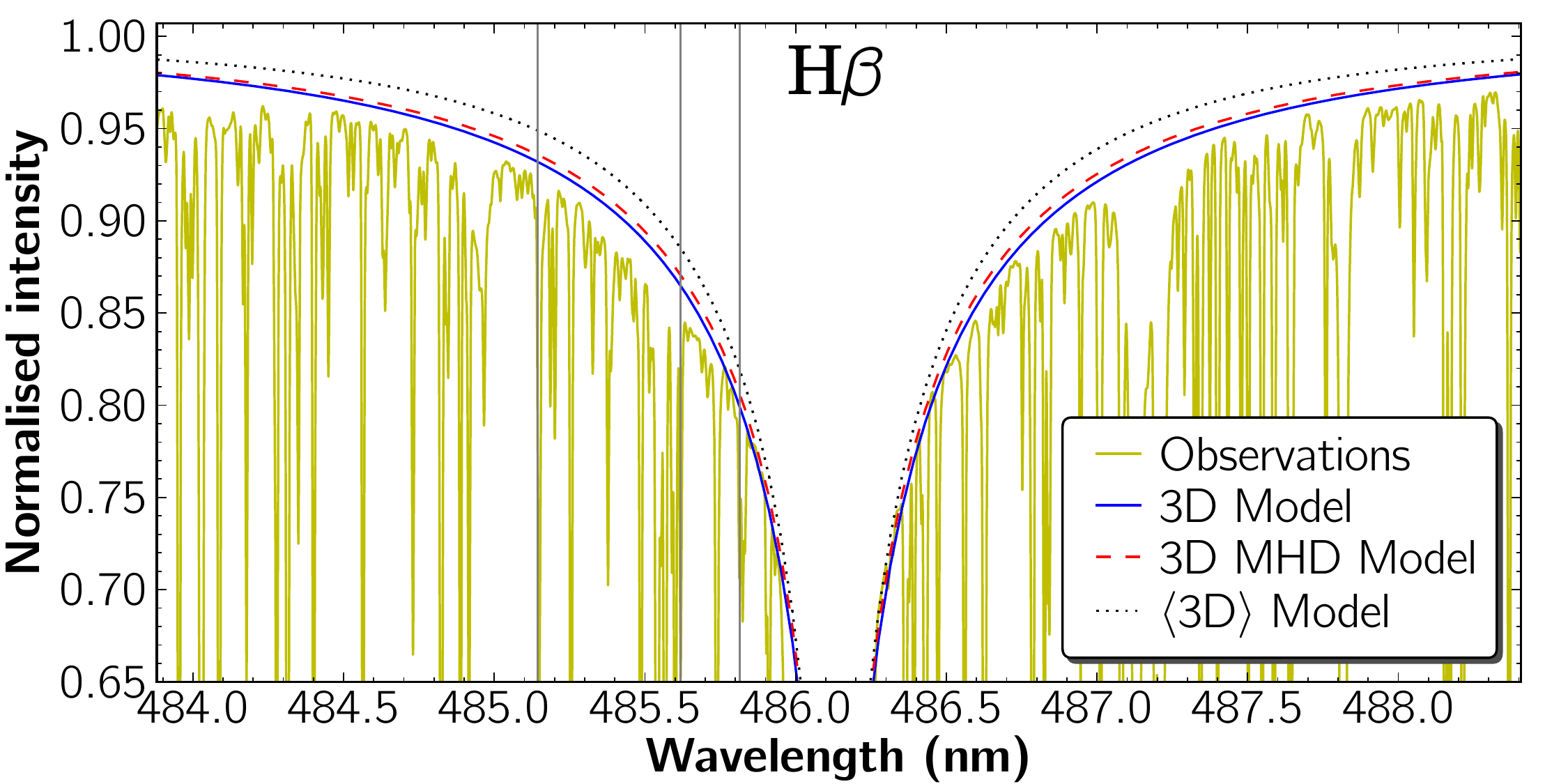} \includegraphics[width=0.49\textwidth]{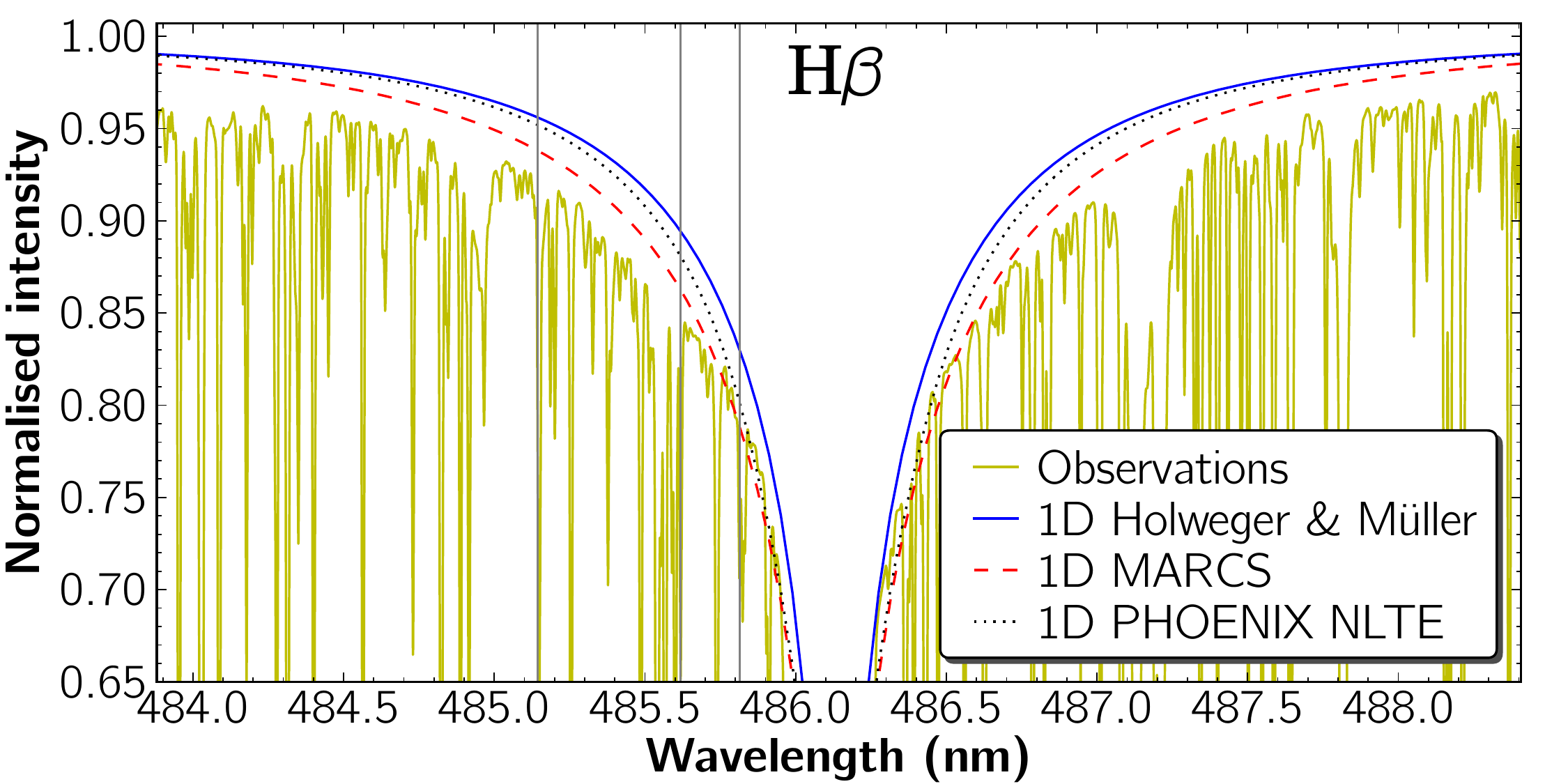}\\
  \includegraphics[width=0.49\textwidth]{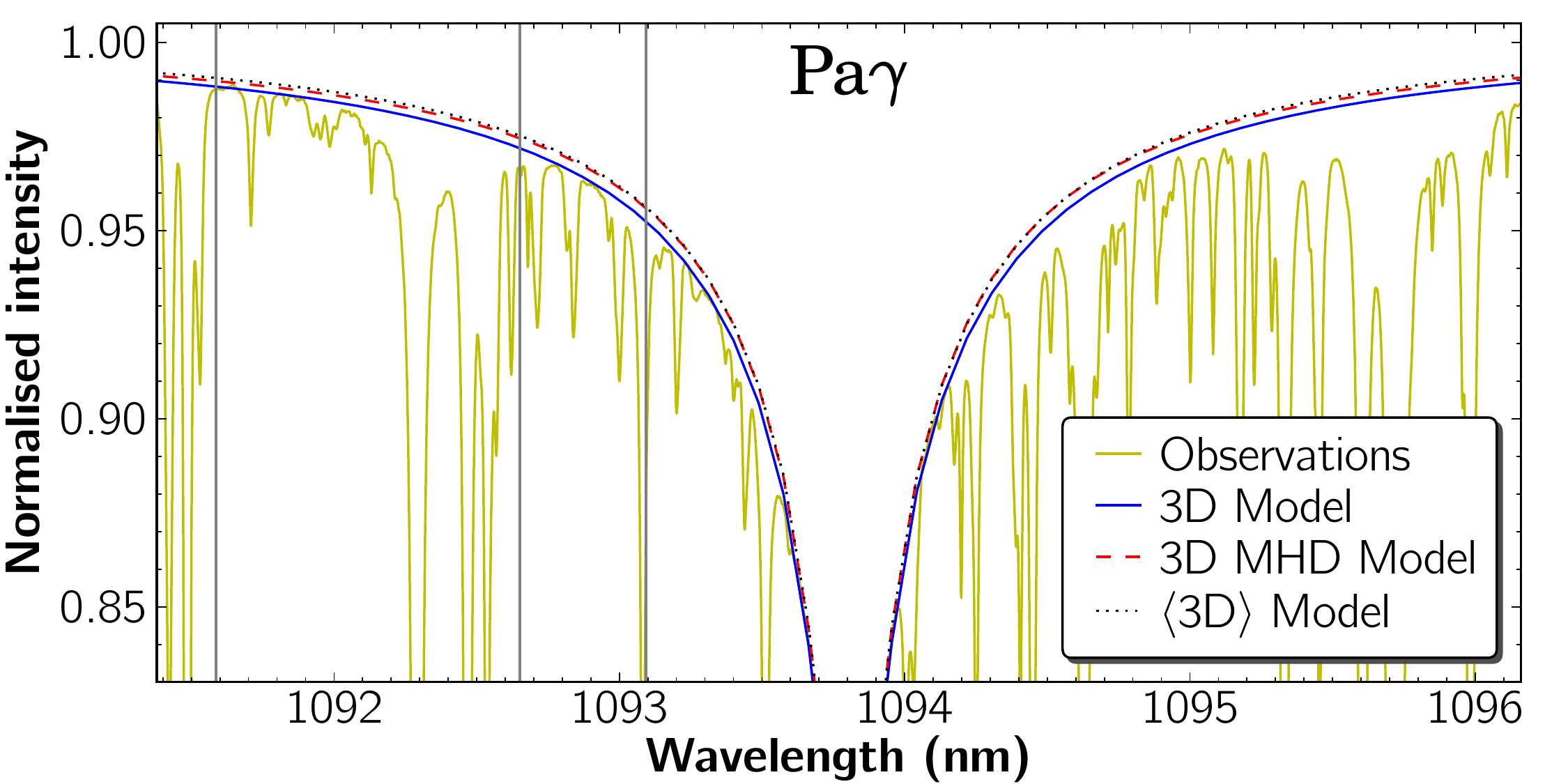} \includegraphics[width=0.49\textwidth]{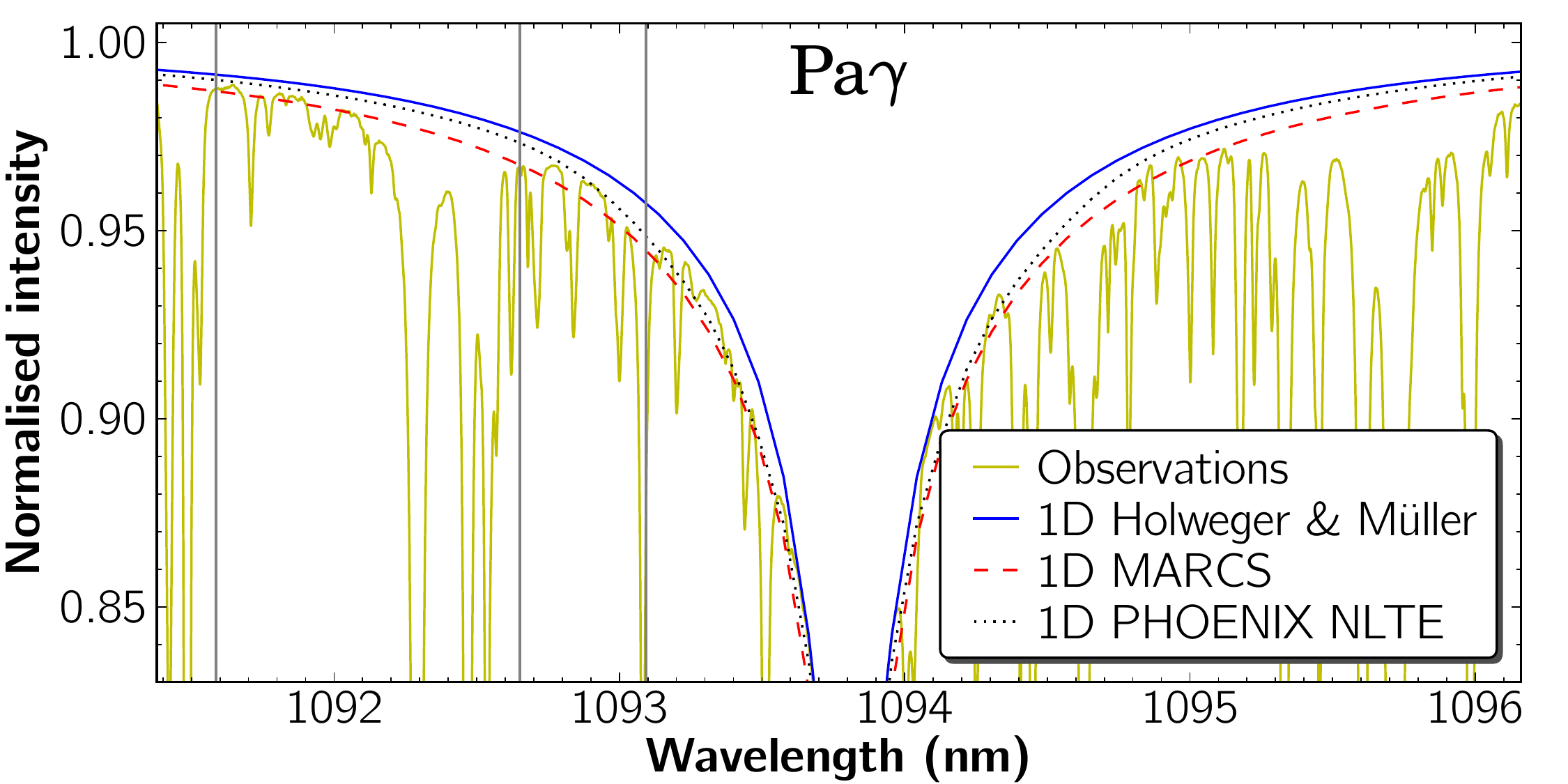}\\
  \caption{Normalised disk-centre intensity profiles for the H lines \hb, \ha, and Pa$\gamma$. Compared with the solar observations of \citet{BraultNeckelFTS}. Synthetic profiles were computed in NLTE. The three vertical lines correspond to the wavelengths $\lambda_1$, $\lambda_2$, and $\lambda_3$ used to make Fig.~\ref{fig:temp_grad}.}
  \label{fig:Hlines_int}  
\end{figure*}

We present the hydrogen line profiles for flux (disk-averaged intensity) and disk-centre intensity. After convergence of the level populations, the flux profiles have been computed using a total of 32 inclined rays (8 $\mu$-angles, 4 $\varphi$-angles); a vertical ray was used for the disk-centre intensity profiles. A rotational velocity of $v_{\mathrm{rot}}=1.8\:\kms$ was used for the disk-integration. For the 1D models a microturbulence of $1.0\:\kms$ and a macroturbulence of $2.5\:\kms$ (consistent with values derived from a sample of Fe\,\textsc{i} lines) were used, although these choices have little effect on the H line wings. 
The hydrogen opacity is calculated using the \texttt{HLINOP} routine (\citealt{BarklemHLINOP}; \citetalias{Barklem2007HNLTE}). This ensures a proper treatment of self-broadening \citep[following][]{Barklem2000} and Stark broadening \citep[following][]{Stehle1999}.

\subsection{Results}

We present the H line profile results in Figs.~\ref{fig:Hlines_flx} and \ref{fig:Hlines_int} for flux and disk-centre intensity respectively. To quantify the differences between the observations and synthetic profiles a $\chi^2$ approach was carried out in the following way. First, regions close to the line-wing continuum were identified in the observations. These regions are indicated in grey in Fig.~\ref{fig:Hlines_flx} and were used for both disk-centre and flux profiles. The differences between observations and synthetic profiles were calculated in these regions. For a tangible quantification of these differences, we have estimated the change in $T_{\mathrm{eff}}$ that each model would need to best match the observations of each line. This was achieved using several MARCS models with \mbox{$5500 \lesssim T_{\mathrm{eff}} \lesssim 6000$} (all with $\log g = 4.44$ and [Fe/H]=0.0), whose hydrogen line profiles were calculated and used to derive an intensity ratio $I(\lambda,T_{\mathrm{eff}})/I(\lambda,T_{\mathrm{eff}}=5777~\mathrm{K})$. This intensity ratio was then multiplied by each synthetic line profile, to obtain the approximate line profile of each model for an arbitrary $T_{\mathrm{eff}}$. This approach is only an approximation, as the variation of the hydrogen lines with $T_{\mathrm{eff}}$ will vary from model to model. Nevertheless, it is good enough for this purpose. Using an optimisation procedure we calculated the $T_{\mathrm{eff}}$ that, for each set of observations, minimises the reduced $\chi^2$, defined as \mbox{$1/N\cdot\sum\left(I_{\mathrm{obs}}-I_{\mathrm{model}}\right)^2/\sigma^2$}, where $N$ is the number of wavelength points minus the degrees of freedom (one, in this case) and $\sigma^2$ the measurement error, which we assume to be constant. These results are shown in Table~\ref{tab:teffs}. Also shown is the reduced $\chi^2$ for the $T_{\mathrm{eff}}$ adjusted line profiles, summed over all the observations, and normalised by the value for the 3D model. This gives a measure of the goodness of the fits.

To help connect the differences between models and observations with the temperature structure of the models we calculated the temperature gradient, defined as:
\begin{equation}
  \label{eq:temp_grad}
  \nabla{}T \equiv \frac{\mathrm{d}\log_{10}T}{\mathrm{d}\log_{10}\tau},
\end{equation}
with the optical depth $\tau$ evaluated at 500~nm. For the disk-centre intensity profiles of H$\alpha$, H$\beta$, and Pa$\gamma$, the temperature gradient was calculated for three regions, corresponding to the formation regions of three wavelength points in the wings of the line profiles. The three wavelength points $\lambda_1$, $\lambda_2$, and $\lambda_3$ are defined as the velocity shifts from the line cores of respectively $-609$, $-316$, and $-196~\kms$. For each wavelength the contribution function is calculated, and $\nabla{}T$ taken as the average of a small depth range around the peak of the contribution function. For the 3D model this is done on a column by column basis, and the final value for $\nabla{}T$ taken as the mean of all the 1D columns in all snapshots, weighted by the continuum intensity of each column. The resulting values of $\nabla{}T$, plotted against line depression of the disk-centre profiles, are shown in Fig.~\ref{fig:temp_grad}. (The wavelength points used are also shown as vertical lines in Fig.~\ref{fig:Hlines_int}.)

\begin{figure} 
  \centering
  \includegraphics[width=0.49\textwidth]{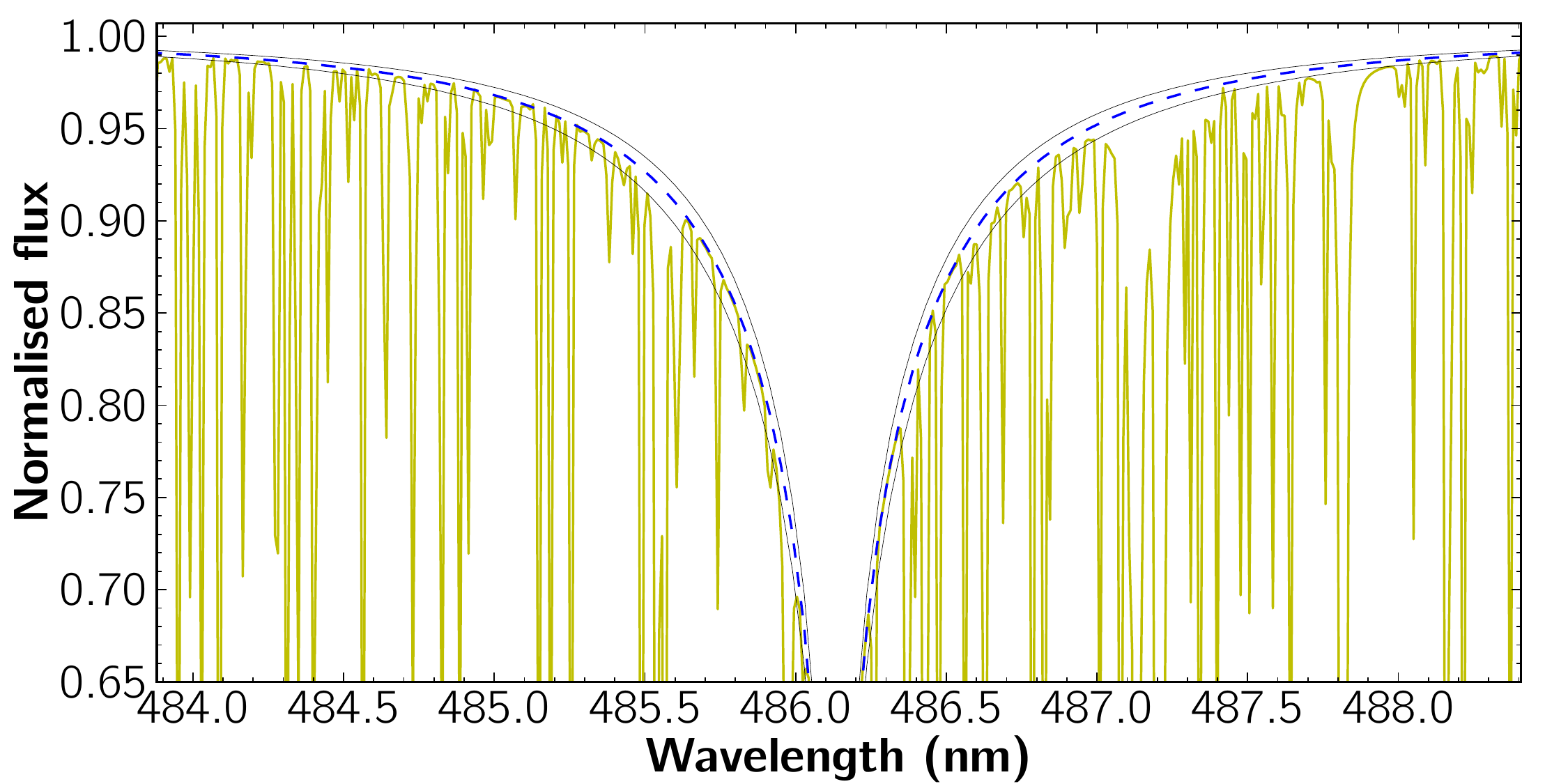} 
  \caption{Effect of multiple blends on the wings of \hb{}. Two synthetic \hb\ flux line profiles of a 1D MARCS model are shown: using only hydrogen (\emph{dashed blue line}), and including 220 other atomic lines (\emph{solid yellow line}). Results for MARCS models with $T_{\mathrm{eff}} + 100$~K and $-100$~K are also shown (\emph{lower and upper thin black lines, respectively}).}
  \label{fig:hbeta_blends}
\end{figure}

\begin{figure} 
  \centering
  \includegraphics[width=0.42\textwidth]{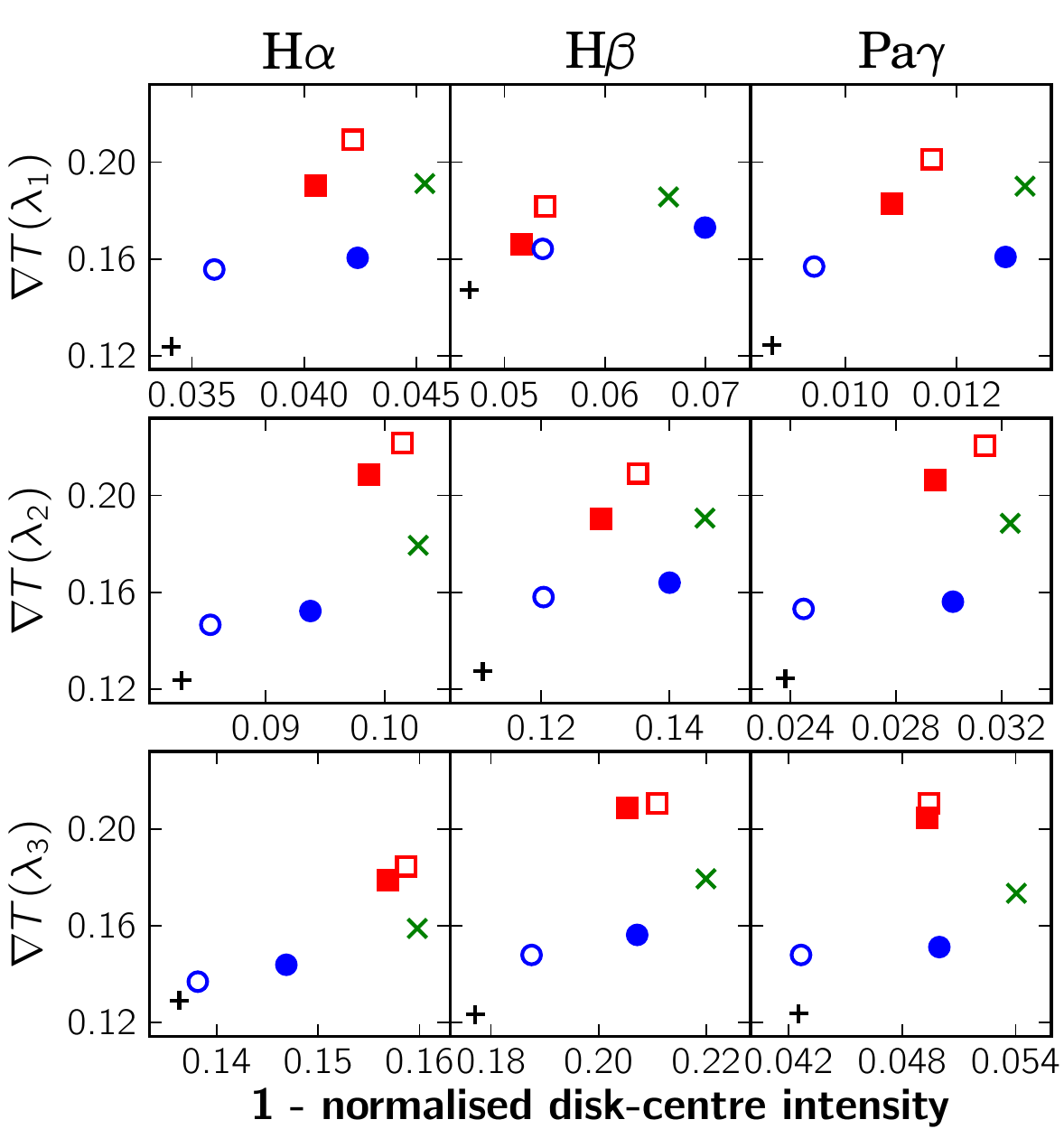} 
  \caption{Temperature gradient at depths corresponding to the formation layers of three wavelengths. Models shown are the 3D (\emph{blue filled circles}), $\langle$3D$\rangle$ (\emph{blue open circles}), Holweger \& M\"uller (\emph{black plus signs}), MARCS (\emph{green crosses}), PHOENIX LTE  (\emph{red filled squares}), and PHOENIX NLTE (\emph{red open squares}).} 
  \label{fig:temp_grad}
\end{figure}

\begin{table*}[ht]
\caption{Estimated effective temperature differences between the models and observations.}
\label{tab:teffs}
\begin{center}
\begin{tabular}{lrrrrrrrrr}
\hline\hline
 & \multicolumn{7}{c}{$\Delta{}T_{\mathrm{eff}}$ [K]} &  & \\
Model  &  \multicolumn{2}{c}{H$\alpha$} & \multicolumn{2}{c}{H$\beta$} &  \multicolumn{1}{c}{Pa$\beta$}  & \multicolumn{2}{c}{Pa$\gamma$} & \multicolumn{1}{c}{$\langle\Delta{}T_{\mathrm{eff}}\rangle{}^\ast$}  & $\chi^2/\chi^2_{\mathrm{3D}}${}${}^\dagger$  \\
  & Flux & $I(\mu=1)$ & Flux & $I(\mu=1)$ & Flux & Flux & $I(\mu=1)$ &  \multicolumn{1}{c}{[K]}  &  \\
\hline
3D Model &              $-50$  & $-53$  & $10$  & $11$  & $66$  & $41$  & $25$  & $\mathbf{7}$  & 1.0\\
3D MHD Model &          $-37$  & $-21$  & $29$  & $49$  & $105$ & $69$  & $66$  & $\mathbf{37}$ & 1.1\\
$\langle$3D$\rangle$ Model &$18$ & $-1$ & $153$ & $136$ & $120$ & $114$ & $69$  & $\mathbf{87}$ & 2.1\\
1D Holweger \& M\"uller & $-10$  & $7$    & $174$ & $196$ & $126$ & $106$ & $82$  & $\mathbf{97}$ & 2.5\\
1D MARCS              & $-98$  & $-156$ & $-9$  & $-30$ & $-17$ & $-7$  & $-74$ & $\mathbf{-56}$& 2.3\\
1D PHOENIX LTE        & $-75$  & $-115$ & $50$  & $81$  & $53$  & $28$  & $-6$  & $\mathbf{2}$  & 3.7\\
1D PHOENIX NLTE       & $-77$  & $-128$ & $37$  & $59$  & $36$  & $14$  & $-27$ & $\mathbf{-12}$& 3.7\\
\hline\hline
\end{tabular}
\end{center}
$^\ast$ $\langle\Delta{}T_{\mathrm{eff}}\rangle$ denotes the simple mean of the seven $\Delta{}T_{\mathrm{eff}}$ values.\\
$^\dagger$ The $\chi^2$ value is the sum of the individual reduced $\chi^2$, normalised by the 3D model results.
\end{table*}

\subsection{Discussion}

\subsubsection{NLTE effects}

The NLTE effects on the hydrogen lines are quantified in Fig.~\ref{fig:hlines_ratios}, where the NLTE/LTE flux ratios are shown. The main consequence of the NLTE effects in the H lines is a deeper core, which is of less interest here since it is little sensitive to the photospheric stratification but rather determined by chromospheric conditions. However, as noted by \citetalias{Barklem2007HNLTE} and shown in Fig.~\ref{fig:hlines_ratios}, there are non-negligible effects on the wings of the Balmer lines, making the wings weaker compared to the LTE case, at least with our particular choice of H collisions for the NLTE calculations \citep[see discussion in][]{Barklem2007HNLTE}). For the Paschen lines the NLTE effects are much smaller, and they cause only a deeper core.

Compared with the results of \citetalias{Barklem2007HNLTE}, we find a weaker NLTE effect in the wings of the Balmer lines. Using a similar MARCS model, the same recipe for the collisional rates, and a similar code for NLTE radiative transfer (MULTI), \citetalias{Barklem2007HNLTE} finds a maximum NLTE excess flux on the wings of H$\alpha$ of about $2.5\%$, whereas we find around $1\%$. The origin of this difference is our inclusion of line-blanketing for photo-ionisation transitions. This additional source of opacity, in particular in the UV, leads to a decrease in the photo-ionisation rates, bringing the level populations closer to LTE. Line-blanketing was not included in the MULTI version 2.2 used by \citetalias{Barklem2007HNLTE}, but is in MULTI version 2.3 and in MULTI3D. If we switch off line-blanketing in MULTI3D, we obtain essentially the same results as \citetalias{Barklem2007HNLTE}.

\subsubsection{Temperature gradient}

The results for the temperature gradient $\nabla{}T$ in Fig.~\ref{fig:temp_grad} show a correlation between the line strength and $\nabla{}T$. Typically, the higher $\nabla{}T$, the stronger the normalised line profiles as one would naively expect. In most cases the Holweger \& M\"uller, $\langle$3D$\rangle$, and PHOENIX models seem to fall on the same linear relation, while the 3D and MARCS models, showing a similar relation, have a lower $\nabla{}T$ for similar line strengths. When compared with the Holweger \& M\"uller, the higher $\nabla{}T$ of the PHOENIX and MARCS models is consistent with their predictions for the wings being much stronger than those of the Holweger \& M\"uller model. However, the $\nabla{}T$ alone is not enough to explain the differences between MARCS and PHOENIX models, with the latter having usually a larger $\nabla{}T$ but a smaller $T_{\mathrm{eff}}$ correction. 

Between the PHOENIX models, the differences in the $\Delta T_{\mathrm{eff}}$ corrections are consistent with what is seen in $\nabla{}T$: the NLTE model has a higher $\nabla{}T$, and consequently also larger $\Delta T_{\mathrm{eff}}$ corrections (positive or negative). Between the Balmer lines, the temperature gradient of the PHOENIX models is lower for \hb. This is a consequence of the abrupt change in temperature gradient that these models show at $\log_{10}\tau_{500}\approx 0.3$ (see Fig.~\ref{fig:ttau}). \hb, being formed deeper, is formed mostly on the flatter side of this knee, whereas \ha\ is mostly formed on the steeper side of the knee. This helps explain why these models predict \ha\ to be too strong, while predicting \hb\ to be too weak. The MARCS model, that does not show this `knee', predicts both Balmer lines to be stronger than the observations.

Perhaps the most interesting departure from the $\nabla{}T$ relation with line strength is the case of the 3D and $\langle$3D$\rangle$ models. These models have a very similar $\nabla{}T$, but the predictions of the $\langle$3D$\rangle$ model are of much weaker lines than the 3D model. For all the tests in the hydrogen lines, the $\langle$3D$\rangle$ is closer to the Holweger \& M\"uller model. This indicates that the spatial and temporal variations of the 3D model are important to describe the shapes and strengths of the hydrogen lines due to the very large non-linearity in the line formation of these high excitation lines. \citet{Ludwig:2009}
reached similar conclusions in their analysis of H lines using 3D CO$^5$BOLD models and stressed that the ``3D effect'' in terms of $T_{\rm eff}$ depends sensitively on the particular adopted mixing length parameters in the 1D model atmospheres. 

\subsubsection{Comparison with observations}

Overall, no single model seems to reproduce the observations perfectly. All of the models tested require different $T_{\mathrm{eff}}$ corrections for different lines. This is particularly true for the Balmer lines, whose $T_{\mathrm{eff}}$ corrections often have opposite signs. Most models predict a too strong H$\alpha$ line, while at the same time H$\beta$ is not strong enough compared to observations. Given the higher number of blending features in \hb\ one wonders if this effect is not caused by a continuum depression because of all the blends, not considered in the hydrogen-only synthetic profiles. To make sure the single-line approximation is valid, we performed spectral synthesis in this region for the 1D MARCS model, including 220 additional atomic lines in the calculations. Data for these lines were extracted from the \textsc{vald} database \citep{VALD1,VALD2}, selected as the strongest lines in this region. The results, shown in Fig.~\ref{fig:hbeta_blends}, indicate that the blends have a negligible effect in lowering the local continuum of the wings of \hb, validating our single-line approximation.

The MARCS model predicts line profiles that are too strong in all cases, requiring a negative $T_{\mathrm{eff}}$ correction. The Holweger \& M\"uller model suffers from the opposite effect: its predictions are considerably weaker than the observations, except for \ha, when they seem to be just about right. Both PHOENIX models seem to behave similary to the MARCS in \ha, but then have mostly positive $T_{\mathrm{eff}}$ corrections for the Paschen lines, a likely consequence of the variations in their temperature gradients in the deepest atmospheric layers, as discussed before. The 3D model predicts the wings of \ha\ to be stronger and the Paschen line wings to be weaker than the observations, but its prediction for \hb\ agrees very well with the observations.

The Balmer line results of the MARCS model can be compared with the LTE results of \citetalias{Barklem2002}. Although \citetalias{Barklem2002} used an earlier MARCS model \citep{Asplund:1997}, slightly different regions for $\chi^2$ with the 1984 Kurucz flux atlas, and possibly slightly different input physics, it is nevertheless relevant to compare our results with theirs. Calculating the LTE MARCS Balmer profiles for the same flux atlas used by \citetalias{Barklem2002}, we derive a $T_{\mathrm{eff}}$ from each Balmer line. For H$\beta$ our determined $T_{\mathrm{eff}}$ is in good agreement with \citetalias{Barklem2002}, but for \ha\ our  $T_{\mathrm{eff}}$ is 75~K lower. While difficult to pinpoint an exact cause, this difference is likely to come from differences in the input physics used to calculate the line profiles.

Except for the $\langle$3D$\rangle$ model, the synthetic profiles of the theoretical models agree better for flux than disk-centre intensity. The biggest difference in fitted $T_{\mathrm{eff}}$ for intensity/flux is found for \ha\, in particular for the MARCS model. This is a hint of shortcomings in the description of the solar temperature profile in the deeper layers of these models, as the disk-centre intensity profiles are formed deeper than the flux profiles. The Holweger \& M\"uller model shows a different trend: its predictions generally agree better for the intensity profiles, and the difference between the corrections for flux and intensity is smaller. It is reassuring to find that the 3D model gives the smallest variation between flux and intensity corrections, again an indication of a realistic temperature profile. 

For all the line profiles considered, the 3D and the PHOENIX LTE and NLTE models give the smallest variation for $\langle\Delta{}T_{\mathrm{eff}}\rangle$, the mean of the $T_{\mathrm{eff}}$ corrections. While a measure of the internal consistency of the models, $\langle\Delta{}T_{\mathrm{eff}}\rangle$ is a simple approximation that does not take into account the shape of the line profiles. With the highest $\chi^2$ values of the models tested, the predictions of the PHOENIX models do not reproduce the shape of the observed line profiles very well, while the 3D model performs the best also in this regard. The Holweger \& M\"uller is the model with the highest $\langle\Delta{}T_{\mathrm{eff}}\rangle$. The difference \mbox{$\langle$3D$\rangle-$3D} is significant for the hydrogen lines: a $\langle{}\Delta T_{\mathrm{eff}}\rangle$ difference of about $80$~K, and a worse $\chi^2$ for the $\langle$3D$\rangle$ model. Overall, the 3D model stands out from the other models. While not perfectly describing the observations, in particular \ha\ and Pa$\beta$, it gives the smallest variation in $T_{\mathrm{eff}}$ between flux and intensity profiles, one of the smallest variations of $T_{\mathrm{eff}}$ among the different lines, and its predicted line shapes agree very well with the observations as evidenced by the lowest $\chi^2$ among the models.

\begin{figure*} 
  \centering
  \includegraphics[width=0.85\textwidth]{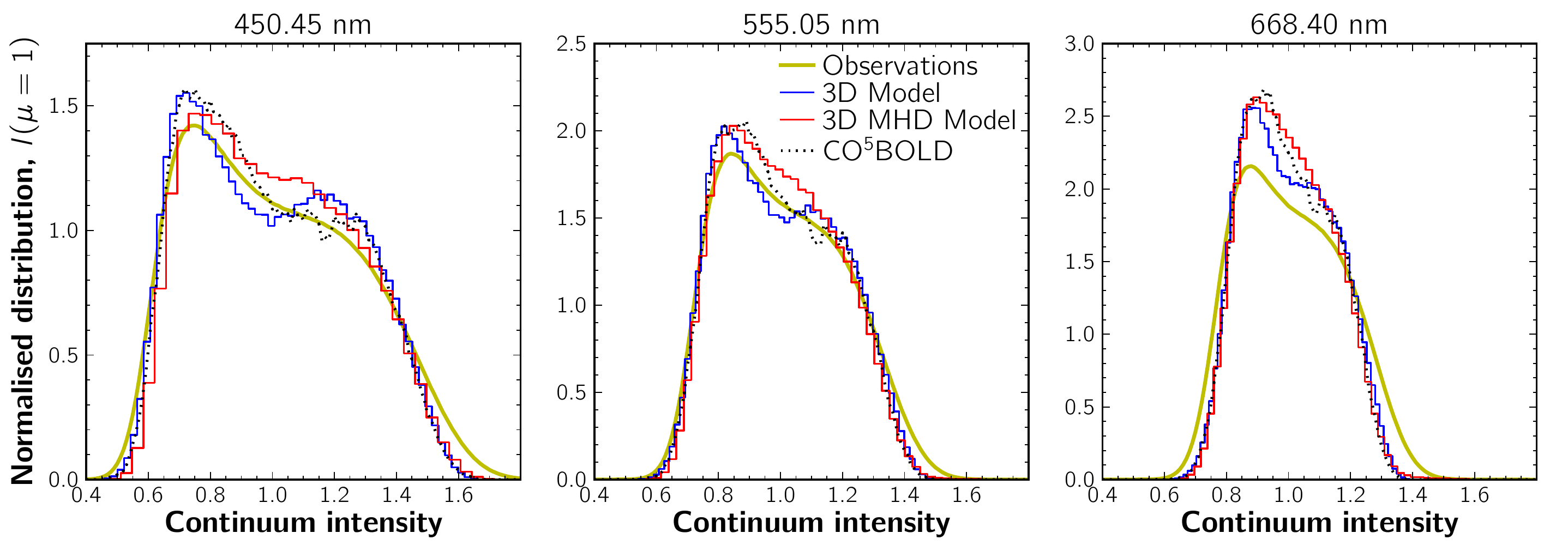} 
  \caption{Continuum intensity distributions for the solar disk-centre for three wavelengths, as a function of the normalised continuum intensity for the 3D model (\emph{thin blue solid line}), the 3D MHD model (\emph{thin red solid line}), compared with observations (\emph{thick yellow line}). Also shown is the prediction from the CO$^5$BOLD 3D solar model (\emph{dotted line}) taken from Fig.~5 of \citetalias{Wedemeyer2009}. The intensity distributions for our 3D models were averaged over all the snapshots.}
  \label{fig:int_dist} 
\end{figure*}

\section{Continuum intensity distribution\label{sec:contint}}

\subsection{Context}

Another relevant diagnostic of our simulations is the continuum intensity
distribution and contrast, $\Delta I_{\rm rms}$, of the granulation.
A comparison of these with observations will reveal how well the
simulations capture the differences in radiative transfer between
the up- and the downflows.

\subsection{Observations}

Because solar observations are made with instruments with a finite resolution and subject to other effects such as straylight, it is difficult to ascertain what the solar intensity distribution and granulation contrast really is. All instrumental effects need to be carefully considered in order to make meaningful comparisons with 3D simulations, but accurately compensating for all optical and straylight effects is a difficult task. 

Several studies have found the observed $\Delta I_{\mathrm{rms}}$ to be lower than that of the 3D simulations even after consideration of atmospheric and instrumental seeing effects \citep[\emph{e.g.}][]{Uitenbroek2007,Kiselman2008}. For the Hinode Solar Optical Telescope \citep[SOT,][]{Tsuneta:2008}, \citet{Danilovic:2008} have characterised in detail the instrumental effects for the Spectro-Polarimeter (SP) instrument, and \citet[][hereafter \citetalias{Wedemeyer2009}]{Wedemeyer2009} have done the same for its Broad Filter Imager (BFI). Both of these studies find that when the instrumental degradation is carefully modelled, the continuum intensity distribution and $\Delta I_{\mathrm{rms}}$ agree very well with the predictions from 3D models. \citetalias{Wedemeyer2009} in particular test several types of 3D models and find an overall good agreement with the space-based observations of Hinode/SOT. 

For our comparison we use the Hinode/BFI observations employed by \citetalias{Wedemeyer2009}. These observations were taken in three BFI channels in three wide-band filtergrams with central wavelengths of 450.45~nm, 555.05~nm, and 668.40~nm. The FWHM of the filter transmission profiles are respectively 0.22~nm, 0.27~nm, and 0.31~nm. The observations were obtained in the period between November 2006 and February 2008. For our comparison, we use the values from \citetalias{Wedemeyer2009} for the observations deconvolved with the instrumental profile (in an attempt to cancel out the image degradation), and compare them with the raw predictions from the simulations (no image degradation applied).

\subsection{Results and discussion}

We compare our 3D models with the observations and the CO$^5$BOLD 3D model results from \citetalias{Wedemeyer2009}. CO$^5$BOLD is independent of the stagger-code we employed, using different numerical methods and atomic physics. Results for the disk-centre intensity distributions at three wavelengths are shown in Fig.~\ref{fig:int_dist}. The $\Delta I_{\mathrm{rms}}$ values for the same wavelengths are given in Table~\ref{tab:irms}. Here $\Delta I_{\mathrm{rms}}$ was calculated in the same way as \citetalias{Wedemeyer2009}, following their equation (1), and averaged in time for all the snapshots considered.

We find that our 3D model reproduces the observations well. Its $\Delta I_{\mathrm{rms}}$ is slightly higher than for the CO$^5$BOLD model, but otherwise results from the two models are very close, which is encouraging. The observed contrast is slightly higher than predicted from our 3D model, probably because (as noted by \citetalias{Wedemeyer2009}) the observations seem to span a wider range of intensities and in particular have a more pronounced `tail' at high intensities. The double peaked structure of the intensity distribution represents the brightness from the inter-granular lanes (highest peak) and the granules (lower peak). Compared to the observations, the 3D model has a pronounced minimum in the distribution between the two peaks, which can be attributed to the lack of magnetic features (such as bright points, ribbons, and other structures).  The 3D MHD model, on the other hand, has a distribution that is closer to the observed, with no local minimum between the two peaks.
This is probably because the bright points and other structures tend to blur the sharp transition from granule to inter-granular lane, and populate the distribution with intensities below the peak attained in the hottest granules. The reduced contrast of the MHD model is consistent with its different stratification and shallower temperature gradient (warmer upper layers), when compared to the 3D model with no magnetic fields. Nevertheless, the 3D MHD model still does not show the high-intensity tail seen in the observations.

\begin{table}[ht]
\caption{Disk-centre $\Delta I_{\mathrm{rms}}$ for the deconvolved observations, our 3D model and the CO$^5$BOLD 3D model.}
\label{tab:irms}
\begin{center}
\begin{tabular}{lrrr}
\hline\hline
$\Delta I_{\mathrm{rms}}$ (\%) &  & $\lambda$ (nm) &  \\
                         & 450.45 & 555.05 & 668.40 \\
\hline
& & & \\
Observations$^{\mathrm{a}}$ & $26.7\pm1.3$ & $19.4\pm1.4$ & $16.6\pm0.7$ \\
3D model      & $25.4\pm0.8$ & $18.6\pm0.6$ & $14.3\pm0.5$ \\
3D MHD model  & $24.1\pm0.8$ & $17.7\pm0.6$ & $13.7\pm0.5$ \\
CO$^5$BOLD$^{\mathrm{a}}$   & $25.0\pm0.1$ & $18.1\pm0.1$ & $13.8\pm0.1$ \\
\hline\hline
\end{tabular}
\end{center}
$^{\mathrm{a}}$ From \citetalias{Wedemeyer2009}.
\end{table}

\section{Fe abundances and line shapes\label{sec:felines}}

\subsection{Context}

The detailed shapes of photospheric absorption lines carry
crucial information about the atmospheric conditions.
The strengths of weak spectral lines mainly reflect the temperature structure in the line-forming 
regions, at least in the framework of LTE, while stronger lines become increasingly
sensitive to the velocity field due to desaturation effects. All observed spectral 
lines show asymmetries 
due to the presence of (anti-)correlations between temperature and velocities in 
the up- and downflows.
The warmer upflows lead to high continuum intensities and blue-shifted, strong line profiles
due to the steep temperature gradients while lines from downflows are red-shifted, weak and
have low continuum intensities; the line strengths are normally heavily biased towards the upflows,
also because of their typically larger area coverage. 
The spatially averaged profiles thus become skewed, 
resulting in lines with typically blue-shifted cores and C-shaped bisectors for 
stars like the Sun \citep[e.g.][ and Fig.~\ref{fig:bisectors}]{Dravins:1982, Asplund2000}. 
Naturally, 1D hydrostatic model atmospheres are unable to explain such line asymmetries
and furthermore require the introduction of two additional free parameters: micro-turbulence to
obtain realistic broadening of partly saturated lines and macro-turbulence to get reasonable
line widths even if the detailed shape of observed lines can never be fully recovered. 
With a self-consistent convective velocity field, 3D models like those employed here do not need
to invoke micro- or macro-turbulence to obtain an excellent agreement with observed line shapes.
Realistic 3D simulations achieve this from first principles, without recourse to adjustable 
parameters \citep{Asplund2000}. 

\citet{Fabbian:2010,Fabbian:2012} have recently investigated the impact of magnetic fields 
on Fe spectral line formation in the quiet Sun, in particular in terms of the inferred solar photospheric Fe 
abundance. Using a series of 3D MHD simulations of varying magnetic field strengths computed with
the Stagger-code, they investigated the 3D LTE line formation of 28 \ion{Fe}{i} lines and found
noticeable differences: lines with small or negligible Zeeman-broadening are still affected by
the different mean temperature structures in 3D models with magnetic fields. For an average vertical
field strength of 10~mT, they found that the derived  \ion{Fe}{i}-based abundance is $\approx 0.05$\,dex higher
than without magnetic fields, the exact effect depending on the particular line in question. 
If this holds true, it would mean that the solar chemical composition presented by
\citet{Asplund2009} would have to be revisited given that it was determined using the same non-magnetic
3D model as we are studying herein. 

\subsection{Observations}

We make use of the solar FTS disk-centre intensity atlas of \citet{BraultNeckelFTS} to measure
the observed line shifts and line bisectors for a sample of \ion{Fe}{i} and \ion{Fe}{ii} lines. 
The FTS atlas is on an absolute radial velocity scale as the relative motion between Sun and Earth
is known precisely and has been corrected for the solar gravitational redshift of 633\,m\,s$^{-1}$ 
(for light intercepted on Earth, \citealt{Lindegren:1999}). The line list stems from
\citet{Asplund2009} and is augmented with additional lines from \citet{Asplund2000}
The necessary laboratory
wavelengths to place the measured line shifts and bisectors on a velocity scale comes from 
\citet{Nave1994} for \ion{Fe}{i} and from Johansson (1998, private communication, see also \citealt{Nave:2012}) for \ion{Fe}{ii}.

\subsection{Results and discussion}

We performed 3D LTE radiative transfer calculations of \ion{Fe}{i} and \ion{Fe}{ii} lines
using the 3D hydrodynamical solar model atmosphere also employed by \citet{Asplund2009} 
as well as a 3D MHD simulation with an average vertical magnetic field of 10~mT from
Thaler et al. (in preparation); we have also ensured that the corresponding 0\,T simulation
of Thaler et al. produces Fe lines indistinguishable from those of the \citet{Asplund2009} model. 
The theoretical disk-centre intensity profiles have been spatially and temporally averaged;
the time sequence corresponds to 45\,min of solar time with snapshots every minute. 
No micro- or macroturbulent broadening entered the 3D line formation calculations but
the resulting averaged line profiles were convolved with a Gaussian corresponding to the
finite resolving power of the \citet{BraultNeckelFTS} solar atlas. 
Polarisation and Zeeman-splitting have not been considered in the radiative transfer calculations.

The solar Fe abundance has been derived from each of our Fe lines using both
the 3D hydrodynamical model and the 3D MHD model. For simplicity, we have
adopted the equivalent widths given in Scott et al. (2013, in preparation), and, when
not available there, in \citet{Asplund2000b}. The magnetic fields can impact the line strengths in two ways: directly via Zeeman-splitting and indirectly via the different atmospheric
stratifications, especially the temperature structure. 
We are not in a position to exactly quantify the former effect but the latter will dominate
for all of our lines due to their small Land\'e factors and relatively short wavelengths  \citet{Fabbian:2012}.
Had we considered Zeeman splitting the derived Fe abundances would be $<0.02$\,dex
{\em smaller} for the few lines with non-negligible Land\'e factors in our line sample for
the 10~mT simulation \citep{Fabbian:2012}.

\begin{figure} 
  \centering
  \includegraphics[width=0.49\textwidth]{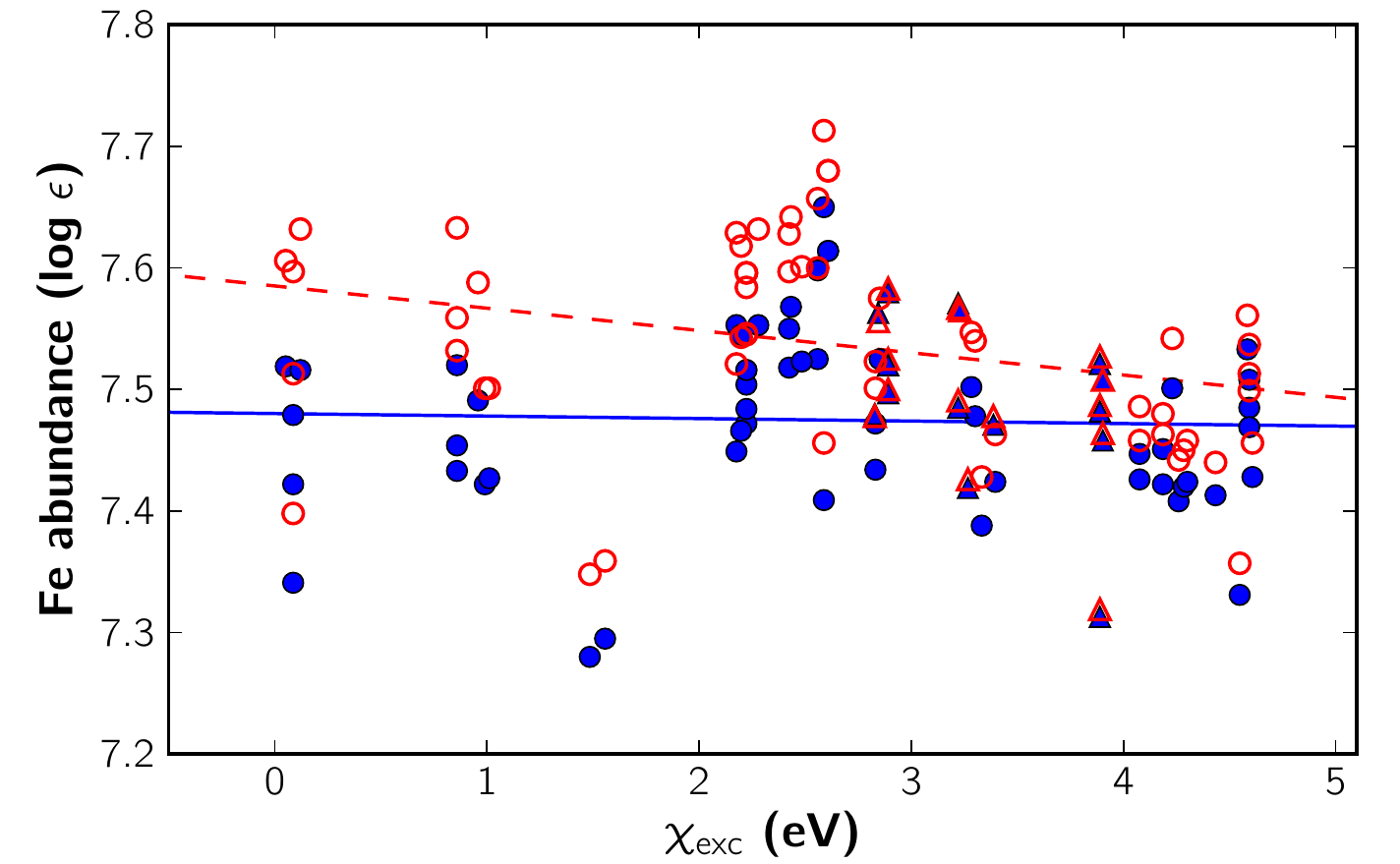}
  \caption{Iron abundances derived from \ion{Fe}{i} lines (circles) and \ion{Fe}{ii} lines (triangles), as a function of excitation potential ($\chi_{\mathrm{exc}}$) for the 3D model (blue, filled symbols) and the 3D MHD model (red, open symbols). The lines show linear fits to the abundance relation for \ion{Fe}{i} lines, for the 3D model (blue solid) and 3D MHD model (red dashed).}
  \label{fig:Fe_abund} 
\end{figure}

\begin{figure} 
  \centering
  \includegraphics[width=0.49\textwidth]{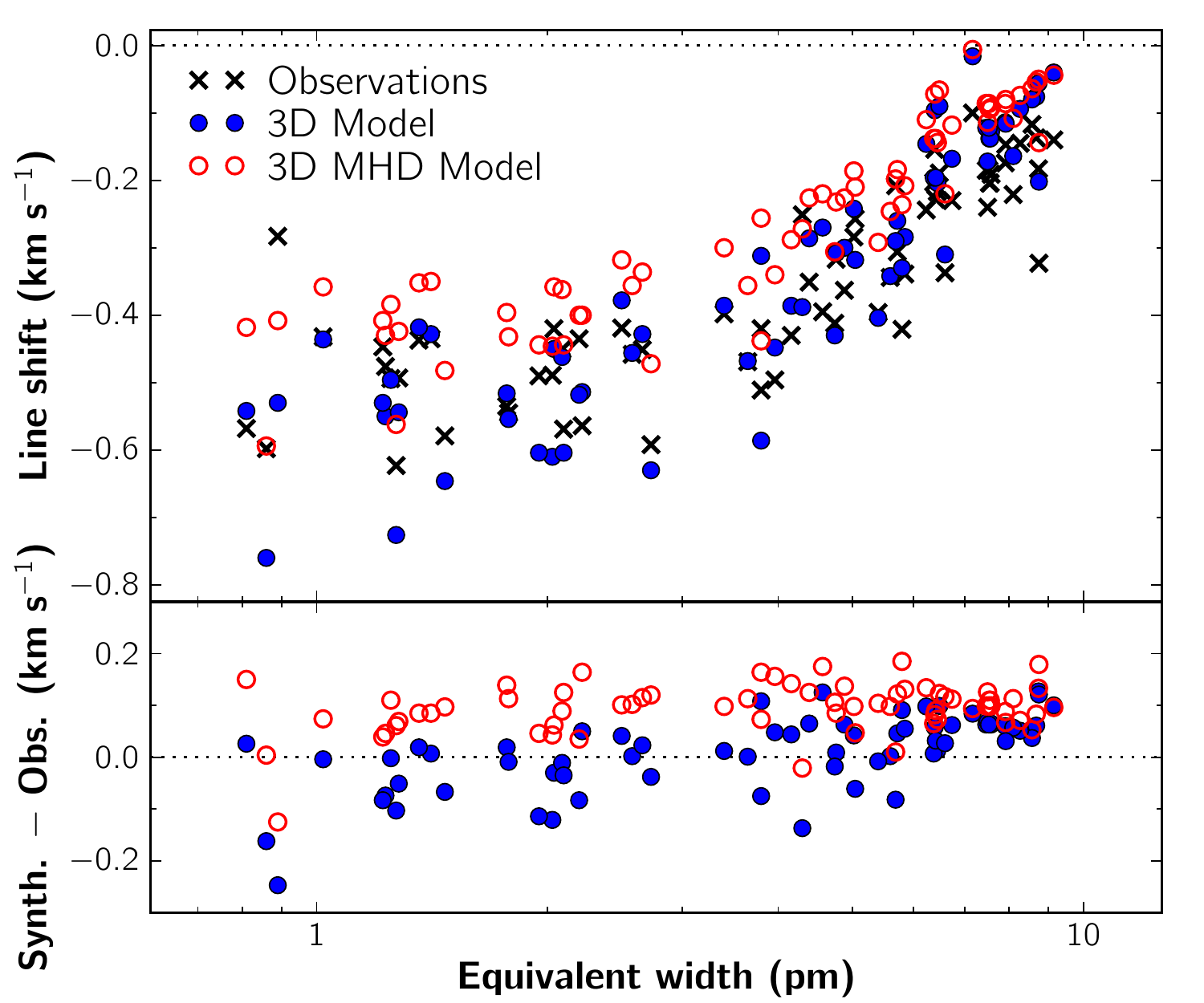}
  \caption{{\em Upper panel:} Observed (crosses) line shifts for a sample of \ion{Fe}{i} and \ion{Fe}{ii} 
  lines for disk-centre intensity as a function of line strength together with predictions from 
  the 3D model (blue, filled circles) and the 3D MHD (100\,Gauss) model (red, open circles). 
  {\em Lower panel:} Differences between predicted and observed line shifts.}
  \label{fig:line_shifts} 
\end{figure}

\begin{figure} 
  \centering
  \includegraphics[width=0.49\textwidth]{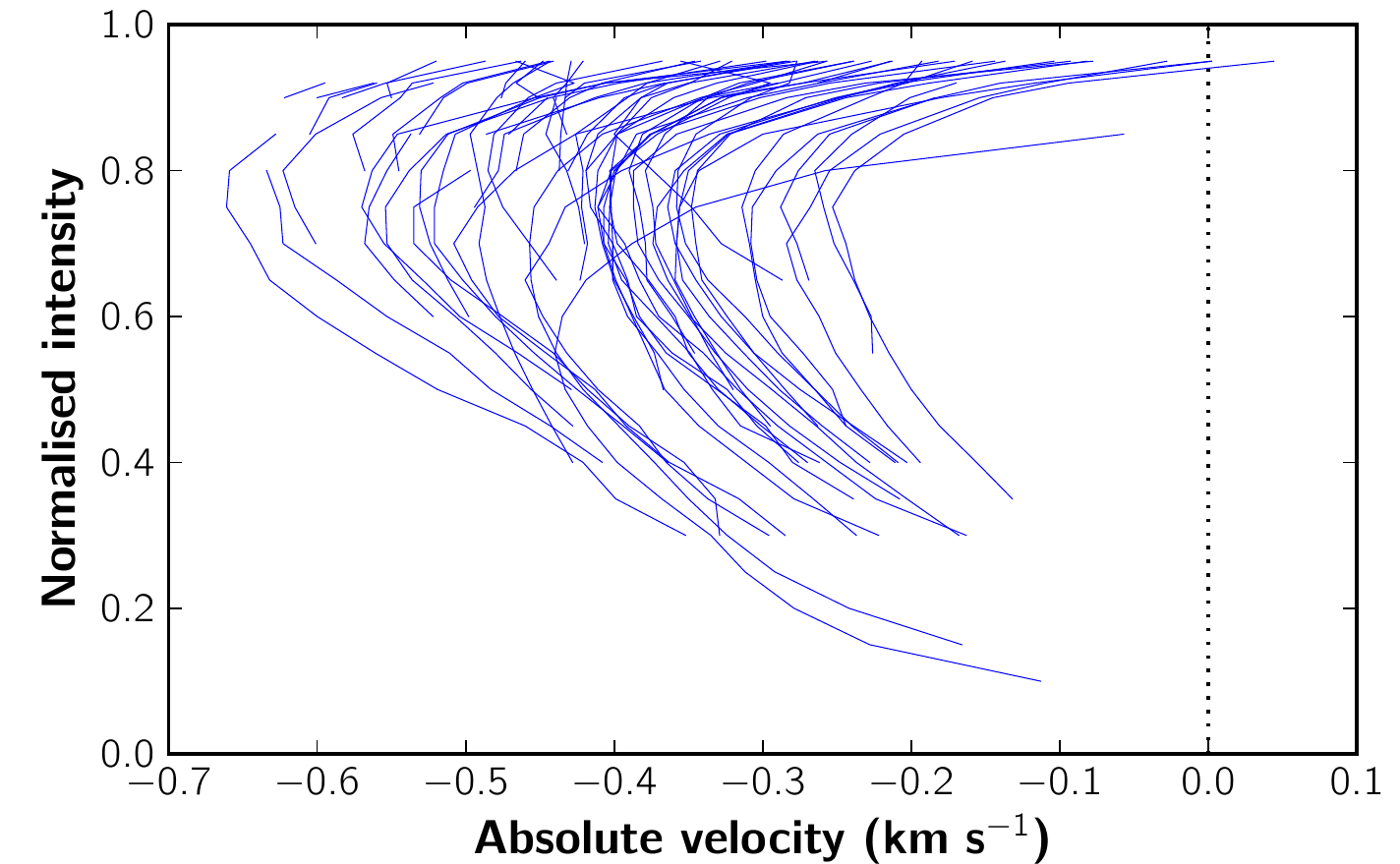}
  \includegraphics[width=0.49\textwidth]{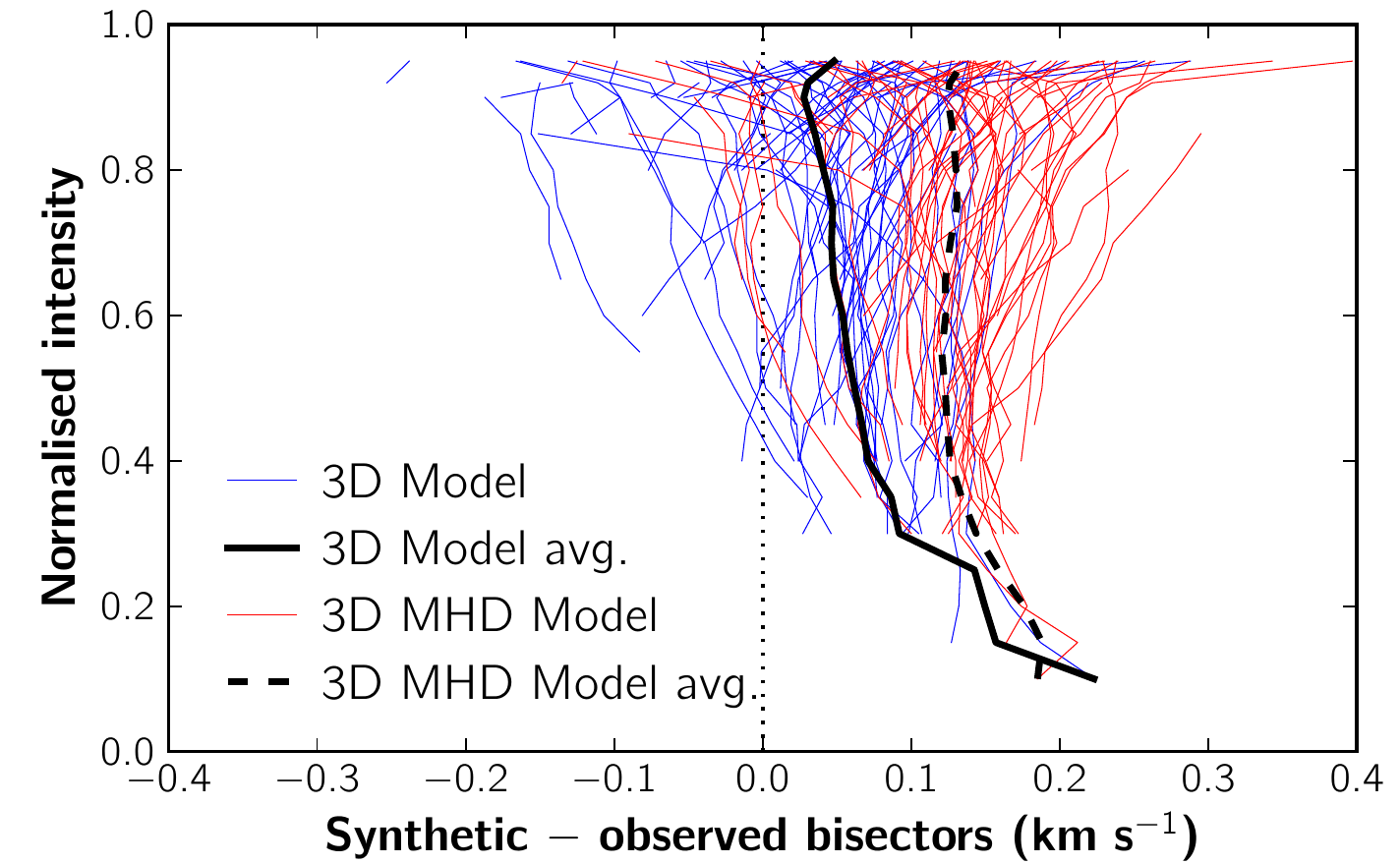}
  \caption{{\em Upper panel:} Observed disk-centre intensity bisectors for a sample of \ion{Fe}{i} and \ion{Fe}{ii} 
  lines. {\em Lower panel:} Differences between predicted and observed bisectors. The thick lines represent the average difference over all bisectors.}
  \label{fig:bisectors} 
\end{figure}

Fig. \ref{fig:Fe_abund} presents a comparison of the Fe abundances from the two 3D models.
Because the MHD model is slightly warmer in the higher atmospheric layers (Fig. \ref{fig:ttau}),
one would expect the inferred \ion{Fe}{i} abundance to be higher than without consideration
of magnetic fields in the convection simulation. Also, the \ion{Fe}{ii} abundances should show very small differences
given their greater formation depths. We find both of these to be true: the MHD mean \ion{Fe}{i} 
abundance is 0.06\,dex higher than without magnetic field, in line with the findings of \citet{Fabbian:2012},
while there is little difference for \ion{Fe}{ii}.  However, as is clear from Fig. \ref{fig:Fe_abund} there
is distinct trend with excitation potential for the 3D LTE \ion{Fe}{i} abundances with the MHD model,
which is not present with the model of \citet{Asplund2009}. It should be noted that this trend cannot
 be removed by departures from LTE because those are positive and more so for low
excitation lines, as demonstrated using the temporally and spatially averaged 
$\langle$3D$\rangle$ model \citep{Bergemann:2012, Lind:2012}.

For both the observed and predicted line profiles, the line centres 
were determined using a cubic spline around the wavelength points with minimum intensity. 
The solar gravitational redshift (of light intercepted at Earth) of 633\,m\,s$^{-1}$ was
subtracted to obtain the observed central wavelengths on an absolute wavelength scale. 
Fig. \ref{fig:line_shifts} shows the observed and predicted \ion{Fe}{i} and \ion{Fe}{ii} line shifts relative to the
adopted laboratory wavelengths of the lines. In line with previous findings, weaker
lines have a more pronounced convective blue-shift due to the larger depths of formation
where convection and the anti-correlations between temperature and velocity are the largest;
the cores of stronger lines become progressively less blue-shifted such that Fe lines
with an equivalent width of $\sim 10$\,pm have nearly vanishing line shifts
\citep{Asplund2000}. 
The agreement between predicted and observed line shifts is very satisfactory for the 
3D hydrodynamical model as demonstrated in the lower panel of Fig. \ref{fig:line_shifts}: 
$30\pm60$\,m\,s$^{-1}$ for our \ion{Fe}{i} lines and
$-50\pm70$\,m\,s$^{-1}$ for \ion{Fe}{ii}. As also found by \citet{Asplund2000} the
stronger Fe lines tend to have slightly underestimated convective blue-shifts;
 \ion{Fe}{i} lines with equivalent widths $<6$\,pm have a mean difference of only $9$\,m\,s$^{-1}$.
 Given the slightly deviating behaviour of two of the weakest lines (\ion{Fe}{i} 669.9\,nm and  \ion{Fe}{ii} 562.7\,nm)
 one could suspect that they are more affected by blends or erroneous laboratory wavelengths
 than the average line.
The predictions from the 3D MHD model have the correct qualitative behaviour but have
systematically too little convective blue-shifts; the mean difference for  \ion{Fe}{i} is 
$100\pm50$\,m\,s$^{-1}$. 

A comparison between observed and predicted line bisectors tell a similar story as 
the line shifts. Solar disk-centre intensity line profiles show a characteristic C-shaped bisector
(weaker lines tend to show only the upper part) with a typical velocity span of 300-600\,m\,s$^{-1}$
with the exact shape depending on the line formation height and temperature/velocity sensitivity
\citep{Asplund2000}. Fig. \ref{fig:bisectors} shows the {\em differences} between the predicted
and observed bisectors for our sample of Fe lines; ideally these differences should manifest themselves
as vertical lines at zero velocity offset. The agreement is very satisfactory for the 3D hydrodynamical model while the bisectors based on the 3D MHD are not sufficiently blue-shifted, in line with the line centre comparison. 

\section{Conclusions\label{sec:conc}}

Realistic solar atmospheres are of paramount importance for our understanding of not just the Sun but also of observations of other stars. The Sun provides an ideal test bench to test the physical ingredients of the models, which if successful can then be applied to other stars with some confidence. A critical requirement for a realistic model is that its thermodynamical quantities such as temperature, density and pressure match those of the real Sun. In this work we have undertaken a systematic study of the temperature structure of several solar models, using several key observational tests: continuum centre-to-limb variation, absolute continuum fluxes, wings of hydrogen lines, and also the intensity fluctuations over the granulation and detailed line shapes and asymmetries.

In all diagnostics we find that the 3D model reproduces the observations very well. This is especially true for the centre-to-limb variations, where its remarkable agreement surpasses even that of the semi-empirical Holweger \& M\"uller model, which was built to fit the centre-to-limb variations. The 3D model also performs very favourably against the absolute continuum fluxes observations. For the hydrogen lines, the 3D model predicts the wings of the \ha\ line to be slightly stronger than the observations, but on the other hand provides a very good agreement for the other lines, and the best overall agreement of all the models tested. In terms of the continuum intensity fluctuations over the solar granulation, it is reassuring to find that the 3D model reproduces the observed intensity distribution and $\Delta I_{\mathrm{rms}}$ well. The 3D model also predicts line shifts and asymmetries that agree very well with observations, which further supports its high degree of realism given the great sensitivity of the exact line shapes on the atmospheric conditions and line formation process. 

In light of the work of \citet{Fabbian:2010,Fabbian:2012}, we also calculated the predictions of a simulation with an average vertical magnetic field of 10~mT (the 3D MHD model). Regarding the Fe line asymmetries, shifts, and abundances, the 3D MHD model agrees slightly less well with observations, suggesting that either the effects of magnetic fields have been overestimated or that it is missing some ingredient that counteracts the consequences of the magnetic fields for the Fe line formation. 
Together with the evidence from the other diagnostics, 
it implies that at this stage there is no justification to prefer the solar abundances
derived from the current generation of 3D MHD solar models over the
3D-based analysis of \citet{Asplund2009}; our results suggest that the 3D MHD Fe abundance corrections advocated by \citet{Fabbian:2010,Fabbian:2012} are over-estimated.

The 1D theoretical models agree well with the observed absolute continuum fluxes, especially the MARCS model. However, both the MARCS and the PHOENIX models predictions for the centre-to-limb variations are consistently below the observations, both in the visible and in the infrared, which we attribute to a too steep temperature gradient. Such 1D hydrostatic models obviously cannot predict any line asymmetries or intensity contrasts. 
We find that the small difference in the temperature structure between the PHOENIX LTE and NLTE models does not translate into any significant difference in our comparison. Their results are very similar. If anything, the NLTE model performs slightly worse against the observational tests. This is likely to result from its somewhat steeper temperature gradient, due to NLTE cooling of the outer layers.

The agreement between the predictions from the 3D model and the observations demonstrates its very high degree of realism in its temperature stratification. Together with its realistic velocity fields and treatment of convection as exemplified by the line asymmetries, it places the 3D modelling in an excellent position to perform chemical abundance studies. It is noteworthy that it greatly outperforms any of the investigated 1D models, both theoretical flavours such as MARCS and PHOENIX and the semi-empirical Holweger \& M\"uller model. There is thus no justification 
to continue to rely on the inferred solar abundances from 1D-based analyses when a significantly improved alternative is available.

\begin{acknowledgements}
We would like to thank Andreas Schweitzer and Peter Hauschildt for kindly providing us with the PHOENIX models, Paul Barklem for the use of his hydrogen collisional data and opacity routines and Sven Wedemeyer-B\"ohm for his observed and synthetic intensity distributions.
TMDP acknowledges financial support from Funda\c c\~ao para a Ci\^encia e Tecnologia (reference number SFRH/BD/21888/2005). This research has been partly funded by a grant from the Australian Research Council (DP0558836).
\end{acknowledgements}

\bibliographystyle{aa}

\end{document}